\documentclass[manuscript,screen]{acmart}
\usepackage{indentfirst}
\usepackage{algorithmic}
\usepackage{bm}
\usepackage{amsmath}

\usepackage{amssymb}
\usepackage{amsthm}
\usepackage{CJK}
\usepackage{color}
\usepackage{graphicx}
\usepackage{indentfirst}
\usepackage{listings}
\usepackage{multicol}
\usepackage{multirow}
\usepackage{diagbox}
\usepackage{subfigure}
\usepackage{threeparttable}
\usepackage{varwidth,array,ragged2e}
\usepackage{listings}
\usepackage{amsfonts}
\usepackage{verbatim}
\usepackage{epstopdf}
\usepackage[ruled,vlined]{algorithm2e}

\usepackage[justification=centering]{caption}

\usepackage{pifont}

\AtBeginDocument{%
  \providecommand\BibTeX{{%
    \normalfont B\kern-0.5em{\scshape i\kern-0.25em b}\kern-0.8em\TeX}}}

\setcopyright{acmcopyright}
\copyrightyear{2020}
\acmYear{2020}

\begin{document}

%%
%% The "title" command has an optional parameter,
%% allowing the author to define a "short title" to be used in page headers.
\title{Locality Sensitive Hash Aggregated Nonlinear Neighbourhood Matrix Factorization for Online Sparse Big Data Analysis}

%%
%% The "author" command and its associated commands are used to define
%% the authors and their affiliations.
%% Of note is the shared affiliation of the first two authors, and the
%% "authornote" and "authornotemark" commands
%% used to denote shared contribution to the research.

\author{Zixuan Li}
\email{zixuanli@hnu.edu.cn}
\affiliation{
  \institution{College of Computer Science and Electronic Engineering, Hunan University}
  \city{Changsha}
  \country{China}
  \postcode{410082}
}

\author{Hao Li}
\email{lihao123@hnu.edu.cn}
\email{H.Li-9@tudelft.nl}
\affiliation{
	\institution{College of Computer Science and Electronic Engineering, Hunan University}
	\city{Changsha}
	\country{China}
	\postcode{410082}
}

\author{Kenli Li*}
\affiliation{%
	\institution{College of Computer Science and Electronic Engineering, Hunan University}
	\city{Changsha}
	\country{China}
}
\email{lkl@hnu.edu.cn, Corresponding Author}

\author{Fan Wu}
\affiliation{%
 \institution{College of Computer Science and Electronic Engineering, Hunan University}
 \city{Changsha}
 \country{China}
}
\email{wufan@hnu.edu.cn}

\author{Lydia Chen}
\affiliation{%
  \institution{Department of Electric Engineering, Mathematics and Computer Science, Distributed Systems, Delft University of Technology}
  \city{Delft}
  \country{Netherlands}
}
\email{lydiayChen@ieee.org}

\author{Keqin Li}
\affiliation{%
  \institution{Department of Computer Science, State University of New York}
  \city{New Paltz}
  \country{USA}
}
\email{lik@newpaltz.edu}

%%
%% By default, the full list of authors will be used in the page
%% headers. Often, this list is too long, and will overlap
%% other information printed in the page headers. This command allows
%% the author to define a more concise list
%% of authors' names for this purpose.
\renewcommand{\shortauthors}{Li, et al.}

%%
%% The abstract is a short summary of the work to be presented in the
%% article.
\begin{abstract}
Matrix factorization (MF) can extract the low-rank features and integrate the information of the data manifold distribution from high-dimensional data,
which can consider the nonlinear neighbourhood information.
Thus, MF has drawn wide attention for low-rank analysis of sparse big data,
e.g., Collaborative Filtering (CF) Recommender Systems, Social Networks, and Quality of Service.
However, the following two problems exist:
1) huge computational overhead for the construction of the Graph Similarity Matrix (GSM), and
2) huge memory overhead for the intermediate GSM.
Therefore, GSM-based MF, e.g., kernel MF, graph regularized MF, etc., cannot be directly applied to the low-rank analysis of sparse big data on cloud and edge platforms.
To solve this intractable problem for sparse big data analysis,
we propose Locality Sensitive Hashing (LSH) aggregated MF (LSH-MF), which can solve the following problems:
1) The proposed probabilistic projection strategy of LSH-MF can avoid the construction of the GSM. Furthermore, LSH-MF can satisfy the requirement for the accurate projection of sparse big data.
2) To run LSH-MF for fine-grained parallelization and online learning on GPUs, we also propose CULSH-MF, which works on CUDA parallelization.
Experimental results show that CULSH-MF can not only reduce the computational time and memory overhead
but also obtain higher accuracy.
Compared with deep learning models,
CULSH-MF can not only save training time but also achieve the same accuracy performance.
\end{abstract}

%%
%% The code below is generated by the tool at http://dl.acm.org/ccs.cfm.
%% Please copy and paste the code instead of the example below.
%%


%%
%% Keywords. The author(s) should pick words that accurately describe
%% the work being presented. Separate the keywords with commas.

%%
%% This command processes the author and affiliation and title
%% information and builds the first part of the formatted document.
\maketitle

\section{Introduction}
In the era of big data,
the data explosion problem has arisen.
Thus,
a real-time and accurate solution to alleviate information overload on industrial platforms is nontrivial \cite{ex113}.
Big data come from human daily needs, i.e., social relationships, medical data and recommendation data from e-commerce companies \cite{ex102}.
Moreover,
due to the large scale and mutability of spatiotemporal data,
sparsity widely exists in big data applications \cite{ex103}.
For accurate big-data processing, representation learning can eliminate redundant information and extract the inherent features of big data,
which makes big-data analysis and processing more accurate and efficient \cite{ex101}.
Furthermore, for sparse data from social networks and recommendation systems,
low-rank representation learning can extract features as latent variables to represent the node and user properties
from the high-dimension space,
which can alleviate the information loss owing to missing data \cite{ex104}.
MF is the state-of-the-art unsupervised representation learning model with the same role as Principal Component Analysis (PCA) and an autoencoder
that can project the high-dimensional space into the low-rank space \cite{ex110}.

Due to its powerful extraction capability for big data,
linear and nonlinear dimensionality reduction is widely used as an emerging low-rank representation learning model \cite{ex111}.
As one of the most popular dimensionality reduction models,
MF can factorize high-dimensional data into two low-rank factor matrices via the constraints of prior knowledge, i.e., distance metrics and regularization items \cite{ex106}.
Then, MF uses the product of the two low-rank matrices to represent the original high-dimension data,
which endows the MF with a strong generalization ability \cite{ex107}.
However,
due to the variety of big data, e.g., multiple attributes of images \cite{ex108}, context-aware text information \cite{ex109}, etc.,
linear MF is not applicable to an environment with hierarchical information; thus,
it should consider the inherent information of big data \cite{ex114}.
Nonlinear MF,
e.g., neural MF \cite{ex115} and the graph for manifold data \cite{ex116} \cite{ex117}, which relies on the construction of the GSM, can mine deep explicit and implicit information.
However, the Deep Learning (DL) model for neural MF needs multilayer parameters to extract inherent variables, which can limit the training speed and
create huge spatial overhead for constructing a GSM; thus, DL cannot be adopted by industrial big data platforms.
Thus, modern industrial platforms are anxious to save parameters in nonlinear MF models \cite{ex181}.

Neighbourhood information for nonlinear MF is an emerging topic \cite{ex118} \cite{ex120}.
The neighbourhood model can strengthen the feature representation by capturing the strong relationship points within the data; and this model is popular in Recommendation Systems, Social Networks,
and Quality of Service (QoS) \cite{ex156} \cite{ex181}.
Handling neighbourhood information is based on several important neighbourhood points that should construct a GSM \cite{ex123}, \cite{ex181}.
However, the use of the GSM should consider the following two problems:
1) the selection and definition of the similarity function should be accurate, and
2) the huge time and spatial overhead caused by the construction of the GSM.
The first problem can be solved by using DL to select the best similarity \cite{ex124}.
However, the huge computational costs make DL unsuitable for cloud-side platforms.
The construction of the GSM takes a huge amount of time and spatial overhead, and its parallelization is difficult.
Due to the quadratically increased spatial costs, the second problem is fatal to real applications using high-dimensional data.
In this case, the approximated strategy is considered to replace the calculation of the GSM.

LSH is a statistical estimation technique that is widely used in high-dimensional data for the Approximate Nearest Neighbourhood (ANN), and
it maps the high-dimensional data to low-dimensional latent space using random projection,
which can simplify the approximated search problem into a matching lookup problem \cite{ex125}.
Due to low time complexity,
LSH has a fast processing capability for high-dimensional data \cite{ex127}.
Furthermore,
LSH has the following drawbacks:
1) the LSH scheme has a slight loss of accuracy;
2) the use of DL can lead to high-precision hashes, but DL is not applicable to cloud-side platforms;
3) online tracking of the hash value for incremental big data;
4) due to information missing,
the similarity between sparse data is not very accurate and should be handled by a specific LSH function.
Thus, it is nontrivial to achieve a reasonable accuracy in less time with fine-grained parallelization for LSH.
Furthermore, with the rapid development of GPU-based cloud-edge computing,
increasingly more vendors will tend to use GPU acceleration \cite{ex128}.
There are three challenges to aggregate LSH with nonlinear MF efficiently to extract the deep features of sparse and high-dimensional data:
1) How can a suitable LSH function be defined to reduce the computation time while ensuring reasonable accuracy?
2) How can LSH be accommodated with the nonlinear neighbourhood MF to achieve low spatial overhead in an online way?
3) How can a GPU and multiple GPUs be used to achieve a faster calculation speed?

This work is proposed to solve the above problems, and the main contributions are presented as follows:

\begin{enumerate}
	\item A novel Stochastic Gradient Descent (SGD) algorithm for MF on a GPU (CUSGD++) is proposed.
	This method can utilize the GPU registers more and disentangle the involved parameters.
	The experimental results show that it achieves the fastest speed compared to the state-of-the-art algorithms.
	\item simLSH is proposed to replace the GSM and accomplish sparse data encoding.
	simLSH can greatly reduce the time and memory overheads and improve the overall approximation accuracy.
	Furthermore, an online method for simLSH is proposed for incremental data.
	\item The proposed CULSH-MF can combine the access optimization on GPU memory of CUSGD++ and the neighbourhood information of simLSH for nonlinear MF.
	Thus, CULSH-MF can complete the training very fast and
	attain an $8000X$ speedup compared to serial algorithms.
	Furthermore, CULSH-MF can achieve a speedup of $2.0X$ compared to CUSGD++.
	Compared with deep learning models, CULSH-MF can achieve the same effect, and CULSH-MF only needs to spend 0.01\% of the training time.
\end{enumerate}

In this work,
related works and preliminary findings are presented in Sections 2 and 3, respectively.
The proposed model for LSH aggregated MF is presented in Section 4.
Experiment results are shown in Section 5.

\section{Rrlated Work}
Owing to the powerful low-rank generalization ability,
MF is widely used in various fields of big data processing, i.e.,
Source Localization \cite{ex129},
Wireless Sensor Networks \cite{ex130},
Network Data Analysis \cite{ex131},
Network Embedding \cite{ex134},
Recommender Systems \cite{ex132}, \cite{ex133},
Hyperspectral Image Classification \cite{ex137} and
Biological Data Analysis \cite{ex135}.
Furthermore, LSH is a powerful hashing tool that can also strengthen the performance of nonlinear dimension reduction, including PCA and MF, for
recommendation \cite{ex166} \cite{ex168}, retrieval \cite{ex167} \cite{ex170} and similarity search \cite{ex169}.
Besides, theoretical research in optimization and machine learning communities, e.g.,
Maximum Margin Matrix Factorization (MMMF) \cite{ex139, ex140},
Nonnegative Matrix Factorization (NMF) \cite{ex141, ex142},
Probabilistic Matrix Factorization (PMF) \cite{ex109, ex143, ex144} and
Weighted Matrix Factorization (WMF)\cite{ex145, ex146, ex147}, also pay considerable attention to MF.
The optimization problem for MF is a classic nonconvex problem \cite{ex136, ex138}.
An alternative minimization strategy, e.g.,
Alternating Least Squares (ALS) \cite{ex148},
SGD \cite{ex149} \cite{ex165}, or
Cyclic Coordinate Descent (CCD) \cite{ex150}, is adopted to solve this nonconvex problem.
An efficient big data processing method requires highly efficient hardware and algorithms.

The rapid development and good performance of GPUs also tend to accelerate basic optimization algorithms
that consider the global memory access, threads and thread block synchronization on a GPU.
Thus, the parallelization processes of related methods on GPUs have unique specialties.
$Tan$ $et$ $al.$ \cite{ex148} proposed cuALS, which parallelizes ALS on a GPU.
$Xie $ $et$ $al.$ \cite{ex149} proposed cuSGD based on data parallelization.
cuSGD \cite{ex149} achieves the goal of acceleration by adopting data parallelization on a GPU,
and it has no load imbalance problem.
$Nisa$ $et$ $al.$ \cite{ex150} optimized the CCD algorithm and proposed the GPU-based CCD++ algorithm.
$Li$ $et$ $al.$ \cite{ex154}, \cite{ex155} proposed CUSNMF based on feature tuple multiplication and summation and CUMSGD based on the elimination of row and column dependencies.
These basic algorithms have good performance on a GPU.
However, scalability is not considered, which results in significant limitations of model compatibility.
Nonlinear MF comprises two components, i.e., a DL model for neural MF \cite{ex153} and a neighbourhood model with GSM for graph MF \cite{ex156} \cite{ex151}.
$He$ $et$ $al.$ \cite{ex152} proposed Neural Collaborative Filtering (NCF) using the DL model, and
this model involves a multilayer neural network that can extract the low-rank feature of MF \cite{ex153}.
The neighbourhood model is often integrated into the algorithm and brings better results \cite{ex156} \cite{ex151}.

The construction of a GSM requires calculating the similarity between high-dimensional points, the choice of
similarity functions play a key role in specific environments, and the selection of the Top-$K$ nearest neighbours from the GSM is time consuming \cite{ex157}.
However, designing an effective similarity function is a difficult task.
Research on training similarities through DL is emerging \cite{ex158}.
However, high-dimensional data cause the computational complexity of DL to dramatically increase.
In order to further optimize the calculation and save space,
pruning strategies and approximation algorithms have been proposed \cite{ex159}.
LSH is such an approximate algorithm based on probability projection \cite{ex179}.
Furthermore, the inverse use of LSH can also achieve the farthest neighbour search \cite{ex165}.
However, most LSH algorithms do not work well in sparse data environments.
minLSH is able to calculate the similarity between sets,
but does not consider the weights of the elements in the set.
Although a considerable amount of work has sought to improve minLSH, this work increases the complexity \cite{ex184}.
simHash \cite{ex179} showed good performance in similar text detection.
LSH can project the feature vectors of similar items to equal hash values with a high probability \cite{ex162}, and
this makes LSH widely used for nearest neighbour searches, fast high-dimensional information searches,
and similarity connections \cite{ex163, ex164}.
Due to the inherent sparsity of big data,
using LSH to construct a GSM to aggregate sparse
MF on a big data platform is nontrivial work.
Furthermore, the accuracy of the low-rank tracking of online learning for incremental big data is a key problem \cite{ex182}.
$Chen$ $et$ $al.$ proposed an online hash for incremental data \cite{ex180}.
However, there is a lack of an online LSH strategy for sparse and online data on parallel and distributed platforms.

\section{Preliminaries}
In this section,
LSH for neighbouring points with closer projective hash values is presented in Section 3.1.
The basic MF model and nonlinear MF with notations are introduced in Section 3.2, and
the related symbols are listed in Table \ref{table21}.

\begin{table}[!htbp]
	\setlength{\abovedisplayskip}{0pt}
	\setlength{\belowdisplayskip}{0pt}
	\renewcommand{\arraystretch}{1.0}
	\caption{Table of symbols.}
	\centering
	\label{table21}
	\tabcolsep1pt
	\begin{tabular}{cl}
		\hline
		\hline
		Symbol \ \ &\ \ Definition\\
		\hline
		$I,J$ \ \ &\ \ Two variable sets with interaction;\\	
		$\textbf{R}$ \ \ &\ \ Input sparse matrix $\in$ $\mathbb{R}^{M\times N}$;\\
		$\widehat{\textbf{R}}$ \ \ &\ \ Low-rank approximated matrix $\in$ $\mathbb{R}^{M\times N}$;\\
		$r_{i,j}$ \ \ &\ \ $(i,j)$th element in matrix $\textbf{R}$;\\
		$\Omega$               \ \ &\ \ The set $(i,j)$ of non-zero value in matrix $\textbf{R}$;\\
		$\Omega_{i}$               \ \ &\ \ The set $j$ of non-zero value in matrix $\textbf{R}$ for variable $I_i$;\\
		$\widehat{\Omega}_{j}$     \ \ &\ \ The set $i$ of non-zero value in matrix $\textbf{R}$ for variable $J_j$;\\
		$\textbf{U}/u_{i}$ \ \ &\ \ Left low-rank feature matrix $\in$ $\mathbb{R}^{M\times F}/$ $i$th row;\\
		$\textbf{V}/v_{j}$ \ \ &\ \ Right low-rank feature matrix $\in$ $\mathbb{R}^{N\times F}/$ $j$th row;\\
		$\mu$ \ \ &\ \  The overall relation between variable set $I$ and variable set $J$;\\
		$b_{i}$ \ \ &\ \  The deviation between variable $I_{i}$ $\in$ $I$ and \textbf{$\mu$};\\
		$\widehat{b}_{j}$ \ \ &\ \ The deviation between variable $J_{j}$ $\in$ $J$ and \textbf{$\mu$};\\
		$\overline{b}_{i,j}$ \ \ &\ \ Overall baseline rating $=\mu + b_{i} + \widehat{b}_{j}$;\\
		\multirow{2}{*}{$n_{j_{1},j_{2}}$} \ \ &\ \ The number of entries in variable set $I$ which have relations with \\
		\ \ &\ \ Variables $\{j_{1}, j_{2}\}$ in variable set $J$;\\
		$\rho_{j_{1},j_{2}}$ \ \ &\ \  Pearson similarity of two variables $\{j_{1}, j_{2}\}$ $\in$ $J$;\\
		$S_{j_{1},j_{2}}$ \ \ &\ \ GSM $\overset{def}{=}\frac{n_{j_{1},j_{2}}}{n_{j_{1},j_{2}}+\lambda_{\rho}}\rho_{j_{1},j_{2}}$;\\
		$R(i)$ \ \ &\ \ The set of variables $\in$ $J$ that are explicitly related to the variable $I_{i}$ $\in$ $I$;\\
		$N(i)$ \ \ &\ \ The set of variables $\in$ $J$ that are implicitly related to the variable $I_{i}$ $\in$ $I$;\\
		$S^{K}(j)$ \ \ &\ \ Top-$K$ Nearest Neighbors variables set of the variable $J_{j}$ $\in$ $J$;\\	
		$J^{K}$ \ \ &\ \ The Top-$K$ Nearest Neighbors Matrix $\textbf{J}^{K}$ $\in$ $\mathbb{R}^{N\times K}$;\\
		$R^{K}(i;j)$ \ \ &\ \ $=R(i)\bigcap S^{K}(j)$;\\
		$N^{K}(i;j)$ \ \ &\ \ $=N(i)\bigcap S^{K}(j)$ ;\\
		\multirow{2}{*}{$\textbf{W}$} \ \ &\ \ Explicit influence matrix $\in$ $\mathbb{R}^{N\times K}$ to represent the degree of explicit\\
		\ \ &\ \ Influence for variable set $J$;\\
		\multirow{2}{*}{$\textbf{C}$} \ \ &\ \ Implicit influence matrix $\in$ $\mathbb{R}^{N\times K}$ to represent the degree of implicit \\
		\ \ &\ \ Influence for variable set $J$;\\
		$w_{j}/w_{j,k_1}$ \ \ &\ \ $j$th Explicit influence vector $\in$ $\mathbb{R}^{K}$ of $\textbf{W}$ $/$ the $k_1$th element of $w_{j}$;\\
		$c_{j}/c_{j,k_2}$ \ \ &\ \ $j$th Implicit influence $\in$ $\mathbb{R}^{K}$ of $\textbf{C}$ $/$ the $k_2$th element of $c_{j}$;\\
		$\overline{I},\overline{J}$ \ \ &\ \ The new variable sets in online learning;\\
		$\widehat{I},\widehat{J}$ \ \ &\ \ Combination of new variable sets and original variable sets in online learning.\\
		\hline
		\hline
	\end{tabular}
\end{table}

\subsection{GSM And LSH}

\begin{definition}[Graph Similarity Matrix (GSM)]
	We assume 2 sets as $I$ $=$ $\{I_{1},\cdots,I_{i},\cdots,I_{M}\}$ and
	$J$ $=$ $\{J_{1},\cdots,J_{j},\cdots,J_{N}\}$.
	Given two variables $\{J_{j_{1}}, J_{j_{2}}\}$ $\in$ $J$ and a similarity function $\mathcal{S}(j_{1}\left|\right|j_{2})$,
	the goal is to construct a weighted fully directed graph $\textbf{G}^{J}$, where each vertex represents a variable in $J$,
	and the weight of each edge represents the similarity of the output vertex to the input vertex calculated by
	$\mathcal{S}(j_{1}\left|\right|j_{2})$.
	The construction of GSM $\textbf{G}^{J}$ should consider the relation between $J$ and $I$.
	The value of $\textbf{G}^{J}_{j_{1},j_{2}}$ relies on $\big\{\{r_{i,j_{1}}|i\in \widehat{\Omega}_{j_{1}}\}, \{r_{i,j_{2}}|i\in \widehat{\Omega}_{j_{2}}\}\big\}$.
\end{definition}

The neighbourhood similarity query for variable set $J$ relies on the GSM $\textbf{G}^{J}$ $\in$ $\mathbb{R}^{N\times N}$ \cite{ex156} \cite{ex181} \cite{ex162}.
The most important problem in the neighbourhood model is to find a set of Top-$K$ similar variables.
For this problem, the Top-$K$ nearest neighbours query is emerging.
\begin{definition}[Top-$K$ Nearest Neighbours]
	
	Given a set of variables $\textbf{S}$, each variable as a vertex constitutes a fully directed graph $\textbf{G}$.
	The goal is to find a subgraph $\textbf{S}^{K}$ where each vertex has $\textbf{K}$ and only $\textbf{K}$ out edges point to the vertices of its Top-$\textbf{K}$ similar variables.	
	
\end{definition}

By querying the GSM, the Top-$K$ nearest neighbours can be obtained.
However, for a large set of variables, the cost of the GSM is huge.
If variable set $J$ has $N$ elements, the computational complexity is $O\big(N(N-1)\big)$.
Furthermore, the overhead for the Top-$K$ nearest neighbours query of a variable $J_{j}$ is $O(2NK-K^{2}+K)$, and
the overhead of Top-$K$ nearest neighbours for the variable set $J$ and the construction of the matrix
$\textbf{J}^{K}$
$\in$ $\mathbb{R}^{N\times K}$ is
$O\big(2N^{2}K-NK^{2}+NK \big)$.
The overall overhead is $O\big(N^{2}(2K+1) + N(K-K^{2}-1) \big)$, and the spatial overhead is $O(NK)$.
Thus, the construction of a GSM using high-dimensional sparse big data is not advisable.
In the context of high-dimensional sparse big data,
the calculation costs of a GSM are squared.
In this case, we need to reduce unnecessary calculations or find an alternative method.
LSH is a probabilistic projection method that projects two similar variables with a high probability to the same hash value
while two dissimilar variables are projected to different hash values with a high probability.
We need to judge the similarity between the two variables and find the Top-$K$ nearest neighbours for each variable.
\begin{definition}[Locality Sensitive Hash (LSH)]
	The LSH function is a hash function that satisfies the following two points:
	\begin{itemize}
		\item For any points $x$ and $y$ in $\mathbb{R}^d$ that	are close to each other, there is a high probability $P_{1}$ that they are mapped to the same hash value $P_{H}[h(x)=h(y)]\geqslant P_{1}$ for $\left|\left|x-y\right|\right|\leqslant R_{1}$; and
		\item For any points $x$ and $y$ in $\mathbb{R}^d$ that are far apart, there is a low probability	$P_{2}$ $<$ $P_{1}$ that they are mapped to the same hash value $P_{H}[h(x)=h(y)]\leqslant P_{2}$ for $\left|\left|x-y\right|\right|\geqslant cR_{1}=R_{2}$.
	\end{itemize}
\end{definition}
The use of LSH has allowed us to reduce the complexity from $O(N^2)$ to $O(N)$.

\begin{figure}[htbp]
	\centering
	\includegraphics[width=6.0in]{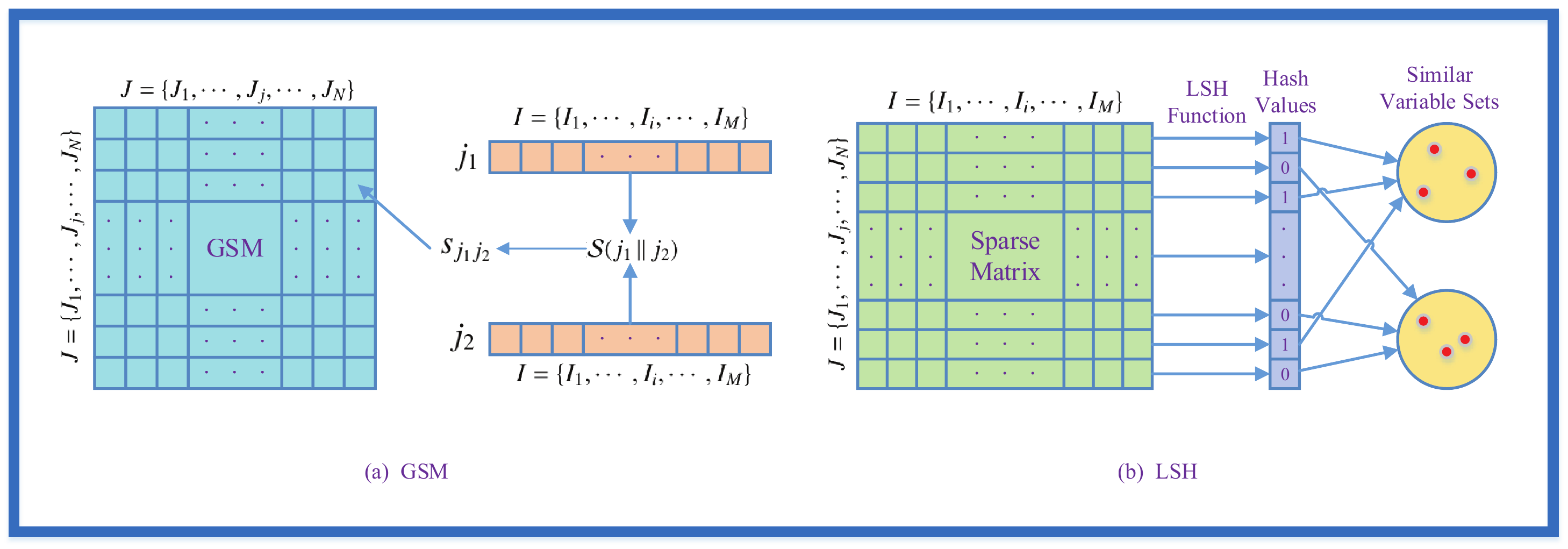}
	\caption{Comparison of computational complexity and space complexity between GSM and LSH}
	\label{gsm_vs_lsh}
\end{figure}

As Fig. \ref{gsm_vs_lsh} shows,
the construction of a GSM requires $O(N^2)$ similarity calculations and consumes $O(N^2)$ space
while the calculation and spatial consumption of LSH is $O(N)$.

LSH can alleviate the problem of huge computational overhead.
However, there are several problems when the LSH is applied to a system with a neighbourhood model:
1) How can a system with a neighbourhood model using LSH obtain the same overall accuracy as the original method?
2) How can the computational model for LSH be incorporated in a big data processing system?
3) How can the system with the LSH model accommodate online learning for incremental data?

\subsection{Nonlinear Matrix Factorization Model}
In big data analysis communities,
representation learning can disentangle the explicit and implicit information behind the data, and the low-rank representation problem is presented as follows.

\begin{definition}[Representation Learning for Sparse Matrix \cite{ex101}]
	Assume a sparse matrix $\textbf{R}$ $\in$ $\mathbb{R}^{M\times N}$ presents the relationship of 2 variable sets
	$\{I, J\}$.
	The value $r_{i,j}$ represents the relation degree of the variables
	$\{I_{i}\}$ in $I$ and $\{J_{j}\}$ in $J$.
	Due to missing information,
	the representation learning task for
	variable $\{I_{i}\}$
	trains
	the feature vector $u_{i}$
	relying on nonzero values $\{r_{i,j}|j\in \Omega_{i}\}$, and
	the representation learning task for
	variable $\{J_{j}\}$
	is to train
	the feature vector $v_{j}$
	relying on nonzero values $\{r_{i,j}| i\in \widehat{\Omega}_{j}\}$.
\end{definition}

\begin{definition}[Sparse Matrix Low-rank Approximation]
	Assume a sparse matrix $\textbf{R}$ $\in$ $\mathbb{R}^{M\times N}$ and a divergence function
	$\mathcal{D}\big(\textbf{R}\|\widehat{\textbf{R}}\big)$ that evaluates the distance between two matrices.
	The purpose of the low-rank approximation is to find an optimal low-rank matrix $\widehat{\textbf{R}}$ and then minimize the divergence.
\end{definition}

MF only involves low-rank feature matrices, and
the feature vectors are used for cluster and social community detection \cite{ex110}.
A sparse matrix has only a few elements that are valuable, and all other elements are zero. Sparse MF is applied to this problem because it factorizes the sparse matrix into two low-rank feature matrices.
In addition, MF model has two limitations:
1) this model is too shallow to capture more affluent features, and
2) this model cannot capture dynamic features.

The approximation value $\widehat{r}_{i,j}$ of the nonlinear matrix factorization model \cite{ex156} is presented as:
\begin{equation}
\begin{aligned}
	\widehat{r}_{i,j}=\underbrace{\overline{b}_{i,j}}_{\textcircled{1}}
	+\underbrace{\left|R^{K}(i;j)\right|^{-\frac{1}{2}}\sum\limits_{J_{j_{1}}\in R^{K}(i;j)}(r_{i,j_{1}}-\overline{b}_{i,j_{1}}){w}_{j,j_{1}}}_{\textcircled{2}}
	+\underbrace{\left|N^{K}(i;j)\right|^{-\frac{1}{2}}\sum\limits_{J_{j_{2}}\in N^{K}(i;j)}{c}_{j,j_{2}}}_{\textcircled{3}}+\underbrace{u_{i}v_{j}^{T}}_{\textcircled{4}}.
\end{aligned}
\label{rr}
\end{equation}

There are $4$ parts in Equation (\ref{rr}), and those parameters can combine the explicit and implicit information of the neighbourhood for nonlinear MF, which are introduced as follows\cite{ex156} \cite{ex181} \cite{ex162}:

\textcircled{1} $\{\mu, b_{i}, \widehat{b}_{j}, \overline{b}_{i,j}\}$:
The baseline score is represented as $\overline{b}_{i,j} = \mu + b_{i} + \widehat{b}_{j}$ for the relation of variable $I_{i}$ $\in$ $I$ and variable $J_{j}$ in set $J$.
Considering that different variables $I_{i}$ $\in$ $I$ have their own different preferences for the entire variable set $J$,
different variables $J_{j}$ $\in$ $J$ have their own different preferences for the entire variable set $I$.
To simplify the description,
suppose \textbf{$\mu$} is the overall relation between variable set $I$ and variable set $J$;
$b_{i}$ represents the deviation between variable $I_{i}$ $\in$ $I$ and \textbf{$\mu$}, which indicates the preference of variable $I_{i}$ to variable set $J$;
and $\widehat{b}_{j}$ represents the deviation between variable $J_{j}$ $\in$ $J$ and \textbf{$\mu$}, which indicates the preference of variable $J_{j}$ to variable set $I$.
A simple case is presented as: $\mu = \sum\limits_{(i,j)\in \Omega} r_{i,j} /|\Omega|$ (the average relation of the known elements),
$b_{i} = \sum\limits_{j\in \Omega_i} r_{i,j} /|\Omega_i| - \mu $
(the difference between the average relation of the known elements in $I_i$ and $\mu$), and
$\widehat{b}_{j} = \sum\limits_{i \in \widehat{\Omega}_{j}} r_{i,j} /|\widehat{\Omega}_{j}| - \mu $
(the difference between the average relation of the known elements in $J_j$ and $\mu$).

$\{n_{j_{1},j_{2}}, S_{j_{1},j_{2}}, S^{K}(j),R(i), R^{K}(i;j), w_{j}\}$:
	Suppose that $J_{j_{1}}$ and $J_{j_{2}}$ are any two variables in $J$, and
	$n_{j_{1},j_{2}} = |\widehat{\Omega}_{j_1} \bigcap \widehat{\Omega}_{j_2}|$
	is the number of variables $\in$ $I$, both of which are related to variables $\{J_{j_{1}}, J_{j_{2}}\}$ $\in$ $J$.
	$\rho_{j_{1},j_{2}}$ is the Pearson similarity for variables $\{J_{j_{1}}, J_{j_{2}}\}$ $\in$ $J$ as a baseline.
	The $(j_{1},j_{2})$th element of GSM is defined as $S_{j_{1},j_{2}}\overset{def}{=}\frac{n_{j_{1},j_{2}}}{n_{j_{1},j_{2}}+\lambda_{\rho}}\rho_{j_{1},j_{2}}$,
	where $\lambda_{\rho}$ is the regularization parameter that adjusts the importance.
	By searching for the GSM, the Top-$K$ nearest neighbours variable set $S^{K}(j)$ of the variable $J_{j}$ $\in$ $J$ can be obtained.
	To retain the generalizability,
	$R(i)$ is denoted as the variable subset of $J$ with explicit relation with variable $I_{i}$ $\in$ $I$,
	which contains all the variables for which ratings by $I_i$ are available.
	If variable $\{J_{j_{1}}\}$ $\in$ $R^{K}(i;j)$ $=$ $R(i)\bigcap S^{K}(j)$,
	variable $I_{i}$ $\in$ $I$ has more explicit relations with variable $J_{j_{1}}$.
	We parameterize the above explicit relations.
	Feature vectors $w_j$ $\in$ $\mathbb{R}^{K}$ are used as the explicit factors for the Top-$K$ nearest neighbours $S^{K}(j)$ of variable $J_{j}$.
	$w_{j,j_{1}}$ is used to represent the information gain that variable $J_{j_{1}}$ $\in$ $R^{K}(i;j)$ explicitly brings to $J_{j}$ $\in$ $J$.
	The closer the basic predicted value $\overline{b}_{i,j}$ is to the true value $r_{i,j}$, the lower the impact received.
Therefore, the residual $(r_{i,j_{1}}-b_{i,j_{1}})$ is used as the coefficient of $w_{j,j_{1}}$.
Combining all $(r_{i,j_{1}}-b_{i,j_{1}}){w}_{j,j_{1}}$, $J_{j_{1}}\in R^{K}(i;j)$ and multiplying the result by a scaling factor $\left|R^{K}(i;j)\right|^{-\frac{1}{2}}$, we obtain $\left|R^{K}(i;j)\right|^{-\frac{1}{2}}\sum\limits_{J_{j_{1}}\in R^{K}(i;j)}(r_{i,j_{1}}-\overline{b}_{i,j_{1}}){w}_{j,j_{1}}$.

\textcircled{3} $\{N(i), N^{K}(i;j), c_{j}\}$:
	To retain the generalizability,
	$N(i)$ is denoted as the variable subset of $J$ with an implicit relation with the variable $I_{i}$ $\in$ $I$,
	and it is not limited to a certain type of implicit data.
	If $\{J_{j_{2}}\}$ $\in$ $N^{K}(i;j)$ $=$ $N(i)\bigcap S^{K}(j)$,
	the variable $I_{i}$ $\in$ $I$ has more implicit relations with variable $J_{j_{2}}$.
	We parameterize the above implicit relations.
	Feature vectors $c_j$ $\in$ $\mathbb{R}^{K}$ are used as the implicit factors for the Top-$K$ nearest neighbours $S^{K}(j)$ of a variable $J_{j}$.
	$c_{j,j_{2}}$ is used to represent the information gain that variable $J_{j_{2}}$ $\in$ $N^{K}(i;j)$ implicitly brings to variable $J_{j}$ $\in$ $J$.
	Combining all ${c}_{j,j_{1}}$, $J_{j_{1}}\in N^{K}(i;j)$ and multiplying the result by a scaling factor $\left|N^{K}(i;j)\right|^{-\frac{1}{2}}$, we obtain $\left|N^{K}(i;j)\right|^{-\frac{1}{2}}\sum\limits_{J_{j_{1}}\in N^{K}(i;j)}{c}_{j,j_{1}}$.

\textcircled{4} $\{u_{i},v_{j}\}$:
	Original MF model.
	$u_{i}$ is the low-rank feature vector for variable $I_{i}$ $\in$ $I$,
	and $v_{j}$ is the low-rank feature vector for variable $J_{j}$ $\in$ $J$.

With the neighbourhood consideration and $L_{2}$ norm constraints for the parameters
$\{\textbf{U},\textbf{V}, \mu, b_{i}, \widehat{b}_{j}, w_{j}, c_{j}\}$, the optimization objective is presented as:

\begin{equation}
\begin{split}
\arg\min\limits_{\textbf{U},\textbf{V}, \mu, b_{i}, \widehat{b}_{j}, w_{j}, c_{j}} \mathcal{D}\big(\textbf{R}\|\widehat{\textbf{R}}\big)
=&\sum\limits_{(i,j)\in\ \Omega}
(r_{i,j}-\widehat{r}_{i,j})^{2}+\lambda_{b}\sum\limits_{i=1}^{M}b_{i}^{2}+\lambda_{\widehat{b}}\sum\limits_{j=1}^{N}\widehat{b}_{j}^{2}\\
&
+\lambda_{w}\sum\limits_{j=1}^{N}\sum\limits_{J_{j_{1}}\in R^{K}(i;j)}w_{j,j_{1}}^2+\lambda_{c}\sum\limits_{j=1}^{N}\sum\limits_{J_{j_{1}}\in N^{K}(i;j)}c_{j,j_{2}}^2
\\ &+\lambda_{u}\sum\limits_{i=1}^{M}\left|\left|{u}_{i}\right|\right|^{2}+\lambda_{v}\sum\limits_{j=1}^{N}\left|\left|{v}_{j}\right|\right|^{2},
\end{split}
\label{loss_r}
\end{equation}
where $\{\lambda_{b}, \lambda_{\widehat{b}}, \lambda_{w}, \lambda_{c}, \lambda_{u}, and \lambda_{v}\}$ are the corresponding regularization parameters.

There are two improvements:
1) the neighbourhood influences are inherent in some big data applications \cite{ex114} \cite{ex118} \cite{ex161}, and
2) the Top-$K$ nearest neighbourhood with explicit and implicit information can replace all queries of neighbourhood points \cite{ex156} \cite{ex181} \cite{ex162}.

\section{Online LSH Aggregated Sparse MF On GPU And Multiple GPUs}
\begin{figure*}[!ht]
	\centering
	\includegraphics[width=6.0in]{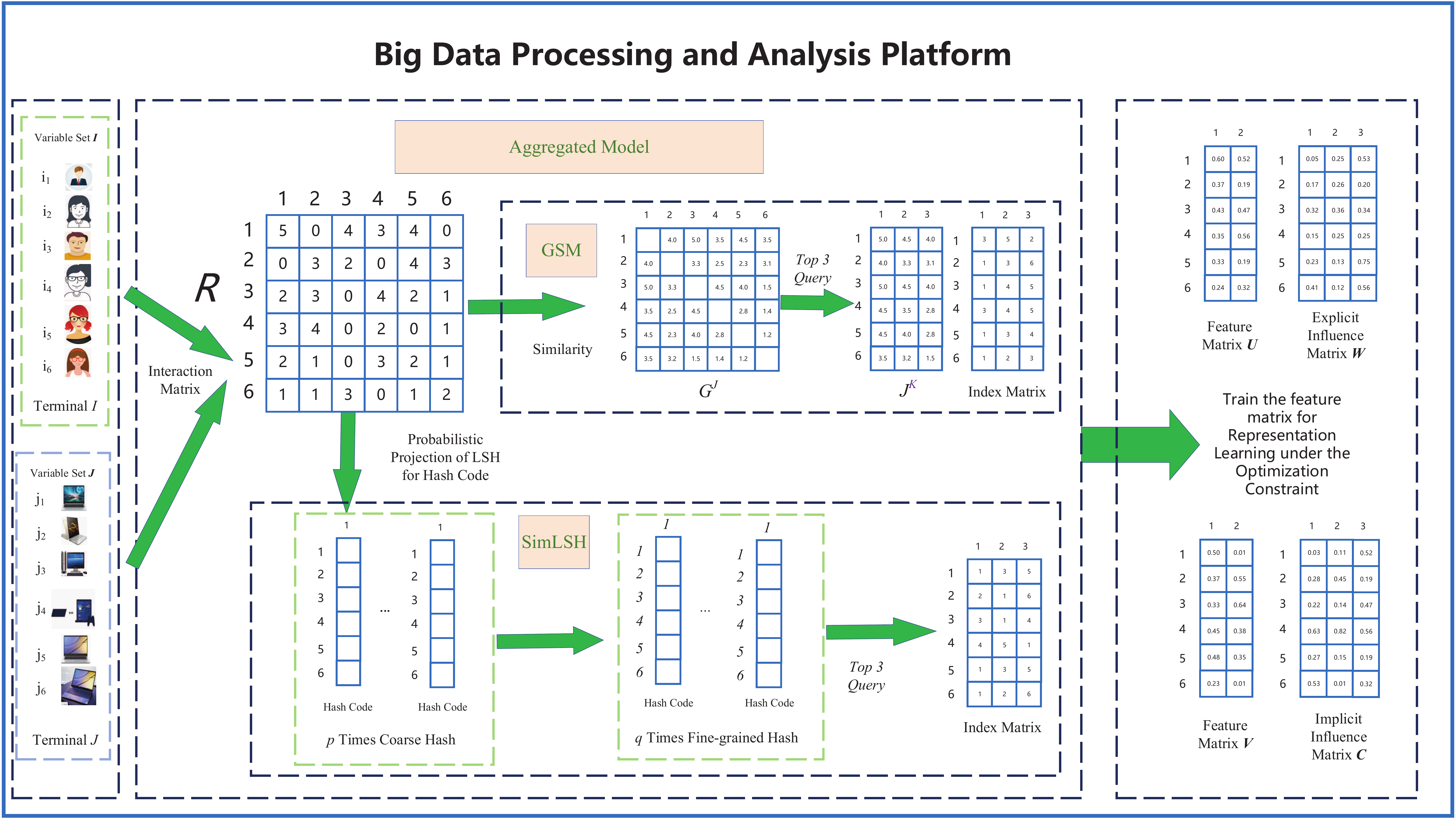}
	\caption{LSH Aggregated Sparse MF on Big Data Analysis Platform}
	\label{Systems_Structure}
\end{figure*}

Fig. \ref{Systems_Structure} illustrates the structure of this work.
First, we consider the interaction value of variable $I_i$ in variable set $I$ and variable $J_j$ in variable set $J$ and generate the interaction matrix $\textbf{R}$ from this.
Second, the original method, which is based on the GSM, can calculate the similarity of every two variables $J_{j_1}$ and $J_{j_2}$ in variable set $J$ to generate a similarity graph $\textbf{G}^{J}$; and
querying $\textbf{G}^{J}$ to obtain the subgraph $\textbf{S}^{K}$ can hold the Top-$K$ nearest neighbours of each variable $J_{j}$ $\in$ $J$.
The difference is that the simLSH method we proposed constructs a hash table through $p$ coarse-grained hashings and $q$ fine-grained hashings.
Then, we obtain the subgraph $\textbf{S}^{K}$ through the hash table.
Finally, we train the feature vectors using the updating rule (\ref{sgd_update}).

As Fig. \ref{Systems_Structure} shows, this work should consider the following three parts:
1) Interaction matrix $\textbf{R}$ of two variable sets $\{I, J\}$, which should consider the incremental data and add the coupling ability of the overall system.
2) The construction of a neighbourhood relationship should reduce the overall space and computational overhead and maintain the overall accuracy.
3) Training the representation feature vectors in a low computational and high accuracy way.
The above objectives guide this section.
In this section,
LSH for sparse big data and CUDA parallelization are presented in Section 4.1;
and then stochastic optimization strategy, CUDA parallelization and multiple GPUs for sparse big data are presented in Section 4.2,
Finally, the online learning solution is presented in Section 4.3.

\subsection{LSH And CUDA Parallelization}
The Top-$K$ nearest neighbours, which relies on the construction of the GSM, is a key step in the nonlinear neighbourhood model.
However, the GSM requires a huge amount of calculations, and the time complexity is $O(N^{2})$ based on the Pearson similarity.
A variety of LSH functions are not friendly to sparse data,

because the accuracy of most distance measures will be greatly reduced.
	This is caused by there being very few positions where the nonzero elements of each vector are the same.
	The Jaccard similarity is suitable for sparse data, and its representative algorithm is minHash\cite{2000Min}; however, this method only considers the existence of the elements and neglects the real value.
In order to solve this problem,
simLSH,
which is inspired by simHash applied to text data, is proposed for sparse dig data projection \cite{ex179}.
This method balances the existence of the elements and the value of the elements and maintains low computational complexity.
simLSH can effectively combine the number of interactions of variable sets $\{I, J\}$ with the degree of interaction, and simLSH can improve the accuracy while reducing the computational complexity.
simLSH is comprised of the following two parts:

1) \textbf{Coding for Sparse Big Data}:

simLSH randomly generates $G$-bits $\{0,1\}$ string $H_i$ for each variable $I_i \in I$,
which is equivalent to a simple hash value.
The hash value $\overline{H}_j$ for each variable $J_j \in J$ that we need is calculated by $H_i$ and $r_{i,j}$, $i \in \overline{\Omega}_{j}$.
Obviously, the hash value $\overline{H}_j$ should also be a $G$-bits $\{0,1\}$ string.
After the hash value $\overline{H}_j$ for variable $J_{j}$ $\in$ $J$ is calculated,
we obtain $\overline{H}_{j,g}$ $\in$ $\overline{H}_{j}$ by accumulating $\Phi(H_{i,g})$ $\cdot$ $\Psi(r_{i,j})$, $i \in \overline{\Omega}_{j}$.
	$\Psi(r_{i,j})$ is a function such that there is a suitable interval between different $r_{i,j}$s,	
	and $\Phi(H_{i,g})$ is a function that maps $H_{i,g}$ from $\{0, 1\}$ to $\{-1,+1\}$.
	Finally,
	$\Upsilon()$ maps the nonnegative value of $\overline{H}_{j,g}$ to $\{1\}$
	and the negative value to $\{0\}$.
Then, the $G$-bit $\{0,1\}$ string $\overline{H}_j$ is obtained.
The entire process of simLSH can be expressed as:

\begin{equation}
\begin{aligned}
	\overline{H}_{j}=\Upsilon\bigg(
	\sum_{i\in \widehat{\Omega}_{j}}\Psi(r_{i,j})\Phi(H_{i})\bigg).
\end{aligned}
\label{item_hash}
\end{equation}

\begin{figure}[htbp]
	\centering
	\includegraphics[width=6.0in]{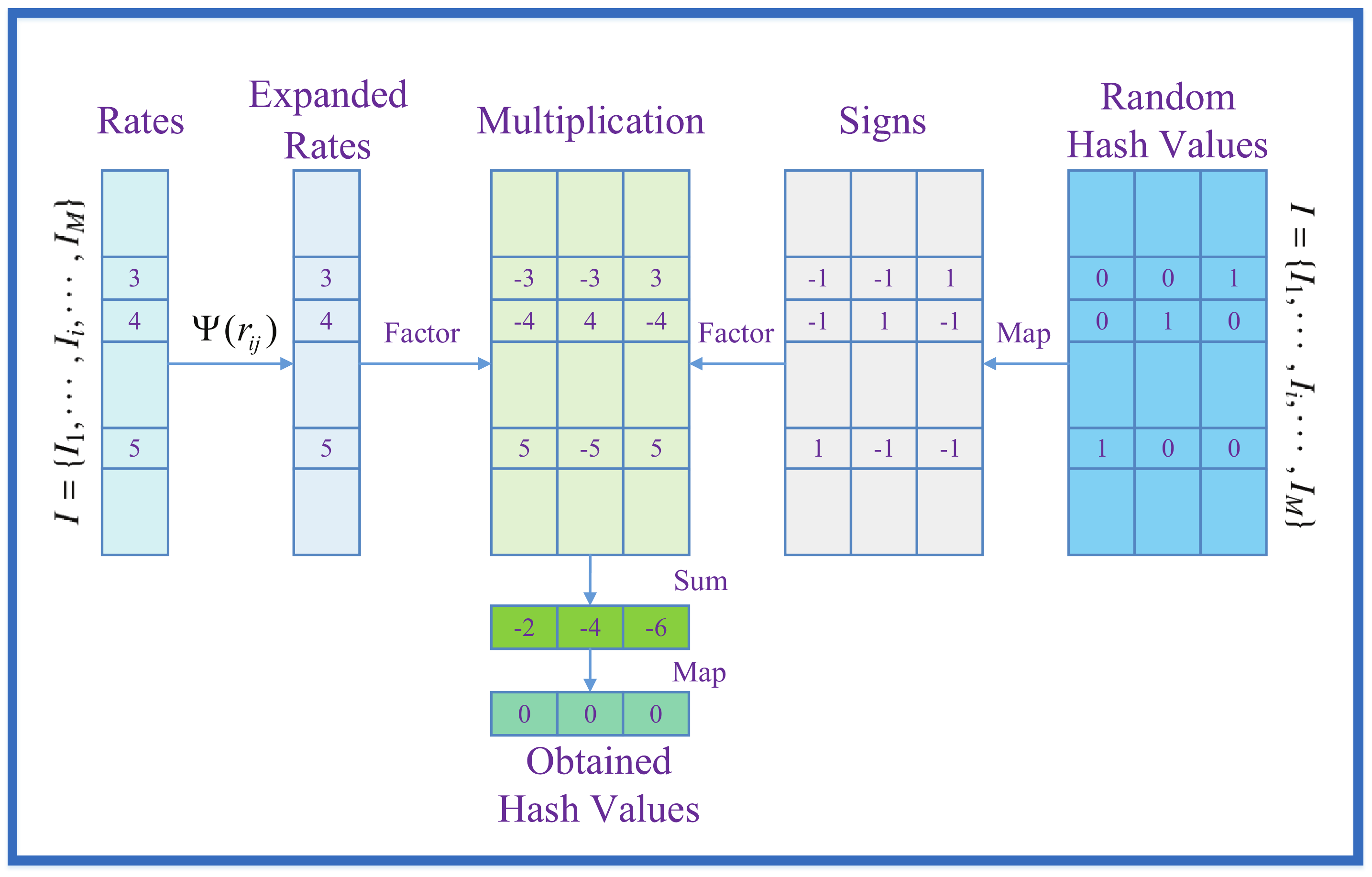}
	\caption{An example of simLSH}
	\label{simlsh_example}
\end{figure}

As Fig. \ref{simlsh_example} shows, variable $J_{j}$ has three relation values $r_{i,j}$ $\{3, 4, 5\}$ with $\{i_{1},i_{2},i_{3}\}$ $\in$ $\overline{\Omega}_{j}$.
When $G$ $=$ $3$,
$\{H_{i_1}, H_{i_2}, H_{i_3}\}$ are randomly assigned to $\{ 001, 010, 100\}$, respectively.
It takes $\Psi(r_{i,j})=r_{i,j}$
by calculating $\big\{(-3-4+5), (-3+4-5), (3-4-5)\big\}$; and then,
the $G$ positions $\{ -2, -4, -6 \}$ of $\overline{H}_{j}$ are obtained, respectively.
Finally, we obtain the $G$-bit $\{0,1\}$ string $\overline{H}_{j}$ $\{0,0,0\}$ by mapping operations.

2) \textbf{Coarse-grained and Fine-grained Hashing}:
LSH is an approximation method to estimate the GSM, but it will achieve accuracy losses when applied to sparse big data.
In this case, simLSH is proposed to speed up the calculations and improve the accuracy.

	Since the maximum probability of two extremely dissimilar variables $\{J_{j_{1}},J_{j_{2}}\}$ with the same hash value is $P_2$,

the mapping of a hash function does not guarantee that the variables $\{J_{j_{1}},J_{j_{2}}\}$ with the same hash value are similar.
In order to alleviate this situation,
the multiple random mapping strategy is considered as follows.
(1) Coarse-grained Hashing:
Similar variables with the same hash values of all mappings are considered.
If $p$ random mappings are conducted, where $p$ $\ll$ $N$,
the probability of two dissimilar variables projected as similar pairs is reduced to at most $P_{2}^{p}$.
Furthermore,
the probability of two similar variables projected as similar pairs is also reduced to at least $P_{1}^{p}$.
Under this condition, many similar variable pairs will be missed.
(2) Fine-grained Hashing:
In this strategy,
as long as at least one of the two variables $\{J_{j_{1}},J_{j_{2}}\}$ projected as similar pairs is subjected to coarse-grained hashing, the similar variable pairs $\{J_{j_{1}},J_{j_{2}}\}$ are selected.
Suppose that $q$ coarse-grained hashings are conducted.
The probability of two similar variables $\{J_{j_{1}},J_{j_{2}}\}$ projected as similar pairs is increase to at least $1-(1-P_1^{p})^{q}$.
By increasing the values of $p$ and $q$, the probability that two similar pairs of variables $\{J_{j_{1}},J_{j_{2}}\}$ are projected as similar pairs is increased.
This method can improve the probability,
and its calculation amount is $p$ $\times$ $q$ times of that of simLSH.
We need to adjust the sizes of $p$ and $q$.
Before the model training,
we only need to perform multiple simLSHs on $N$ variables to find similar variable pairs,
which can reduce the computational complexity to $O(N)$.
Even if you use $p \times q$ simLSHs to increase the probability,
the computational complexity is only $p \times q \times N$, and $p \times q \times N$ is much smaller than $N^2$.

Our goal is to find the Top-$K$ nearest neighbours for each variable $J_{j}$ $\in$ $J$.
simLSH does not directly obtain the Top-$K$ nearest neighbours for $J_{j}$.
It is accomplished by searching for other variables with the same hash value in the hash table.
We use the coarse-grained and fine-grained hashing of simLSH and select the $K$ most frequent variables
$\{J_{1},\cdots,J_{K}\}$ $\in$ $J$ in the hash table of variable $J_{j}$ and make a random supplement if the number is less than $K$.
On the CUDA platform, each thread block for simLSH (CULSH) manages a variable $J_{j}$.
CULSH is described in Algorithm \ref{culsh} as follows:
(1) Lines $1-9$:
	The calculation of simLSH with coarse-grained hashing and fine-grained hashing.
In lines 3-5, calculate the hash value $\overline{H}_{j}$ for variable $J_{j}$ $\in$ $J$ in parallel and save it,
and this only consumes a small amount of memory.
(2) Lines $10-12$:
Search the Top-$K$ nearest neighbours $\{J_{j_{1}}, \cdots, J_{j_{K}}\}$ of variable $J_{j}$ $\in$ $J$ according to hash value $\overline{H}_{j}$ of variable $J_{j}$ $\in$ $J$.

\begin{algorithm}[htbp]
	\caption{CULSH}
	\label{culsh}
	\vspace{.1cm}
	$\textbf{Input}$: Sparse matrix $\textbf{R}$ of variable sets $\{I, J\}$, Random Hash values $H_i$.\\
	$\textbf{Output}$: The Top-$K$ Nearest Neighbors Matrix $\textbf{J}^{K}$ $\in$ $\mathbb{R}^{N\times K}$. Each row represents the Top-$K$ Nearest Neighbors of a variable $J_{j}$ $\in$ $J$.\\
	\begin{algorithmic}[1]	
		\FOR{$\textbf{(Fine-grained Hashing)}$: $q$ times Coarse-grained Hashing}
		\FOR{$\textbf{(Coarse-grained Hashing)}$: $p$ times simLSH}
		\FOR{$\textbf{(parallel)}$: Variables $J_j$ $\in$ $J$ are evenly assigned to thread blocks $\bigl\{TB_{tb\_idx}|tb\_idx\in\{1,\cdots,TB\}\bigl\}$}
		\STATE Calculate the hash value $\overline{H}_{j}$ by equation (\ref{item_hash}) for variable $J_{j}$ $\in$ $J$.
		\ENDFOR
		\ENDFOR
		\STATE Count the similar variable pairs with the same hash value in $p$ times simLSH.
		\ENDFOR
		\STATE Count the similar variable pairs that appear one or more times in $q$ coarse-grained hashings.
		\FOR{$\textbf{(parallel)}$: Variables $J_j$ $\in$ $J$ are evenly assigned to thread blocks $\bigl\{TB_{tb\_idx}|tb\_idx\in\{1,\cdots,TB\}\bigl\}$}
		\STATE Search the Top-$K$ nearest neighbours $\{J_{j_{1}}, \cdots, J_{j_{K}}\}$ of the variable $J_{j}$ $\in$ $J$.
		\ENDFOR
	\end{algorithmic}
\end{algorithm}

\subsection{Stochastic Optimization Strategy And CUDA Parallelization On GPUs And Multiple GPUs}
\begin{figure}[htbp]
	\centering
	\includegraphics[width=6.0in]{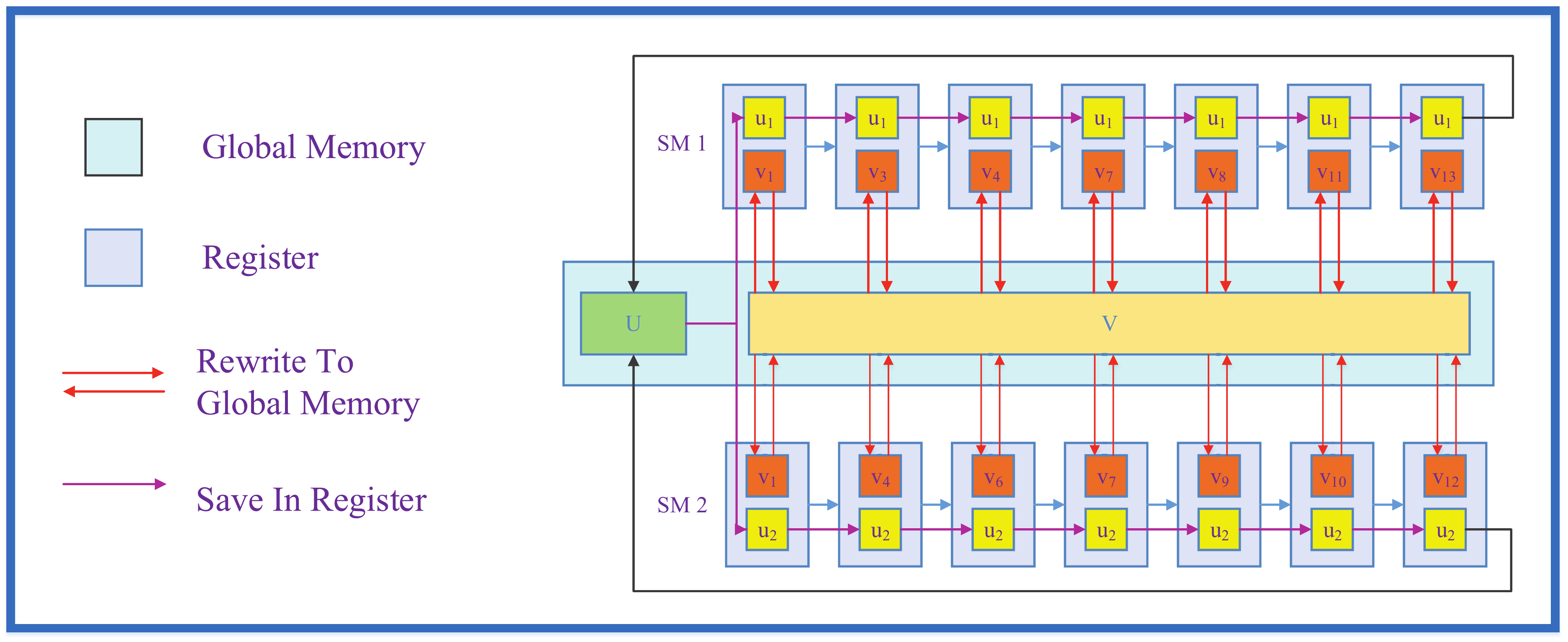}
	\caption{Memory Access Model of CUSGD++}
	\label{memory_access}
\end{figure}

The basic optimization objective (\ref{loss_r}) involves 6 tangled parameters $\{\textbf{U},\textbf{V}, b_{i}, \widehat{b}_{j}, w_{j}, c_{j}\}$.
The state-of-the-art parallel strategy of SGD in \cite{ex149} \cite{ex165} cannot disentangle the involved parameters.
Due to the entanglement of the parameters,
the optimization objective (\ref{loss_r}) is nonconvex, and alternative minimization is adopted \cite{ex136} \cite{ex138} \cite{ex148} \cite{ex150}, which can disentangle the involved parameters as follows:
\begin{small}
	\begin{equation}
	\begin{aligned}
	\left\{
	\begin{aligned}
	&\arg\min\limits_{u_{i}} \sum\limits_{j\in \Omega_{i}}\big(r_{i,j}-\widehat{r}_{i,j}\big)^{2}+
	\lambda_{u}\sum\limits_{i=1}^{M}\left|\left|{u}_{i}\right|\right|^{2};\\
	&\arg\min\limits_{v_{j}} \sum\limits_{i\in \widehat{\Omega}_{j}}\big(r_{i,j}-\widehat{r}_{i,j}\big)^{2}+
	\lambda_{v}\sum\limits_{j=1}^{N}\left|\left|{v}_{j}\right|\right|^{2};\\	
	&\arg\min\limits_{b_{i}} \sum\limits_{j\in \Omega_{i}}\big(r_{i,j}-\widehat{r}_{i,j}\big)^{2}+
	\lambda_{b}\sum\limits_{i=1}^{M}b_{i}^{2};
	\\
	&\arg\min\limits_{\widehat{b}_{j}} \sum\limits_{i\in \widehat{\Omega}_{j}}\big(r_{i,j}-\widehat{r}_{i,j}\big)^{2}+
	\lambda_{\widehat{b}}\sum\limits_{j=1}^{N}\widehat{b}_{j}^{2};
	\\	
	&\arg\min\limits_{w_{j,j_{1}}} \sum\limits_{i\in \widehat{\Omega}_{j}}\big(r_{i,j}-\widehat{r}_{i,j}\big)^{2}+
	\lambda_{w}\sum\limits_{J_{j_{1}}\in R^{K}(i;j)}w_{j,j_{1}}^2;
	\\
	&\arg\min\limits_{c_{j,j_{2}}} \sum\limits_{i\in \widehat{\Omega}_{j}}\big(r_{i,j}-\widehat{r}_{i,j}\big)^{2}+
	\lambda_{c}\sum\limits_{J_{j_{2}}\in N^{K}(i;j)}c_{j,j_{2}}^2.
	\\
	\end{aligned}
	\right.\\
	\end{aligned}
	\label{easy_alternative}
	\end{equation}
\end{small}
SGD is a powerful optimization strategy for large-scale optimization problems \cite{ex138} \cite{ex148}.
Using SGD to solve the optimization problem (\ref{easy_alternative}) is presented as:
\begin{small}
	\begin{equation}
	\begin{aligned}
	\left\{
	\begin{aligned}
	b_{i}&\leftarrow b_{i}+\gamma_{b_{i}}\bigg(e_{i,j}-\lambda_{b}b_{i}\bigg);\\
	\widehat{b}_{j}&\leftarrow \widehat{b}_{j}+\gamma_{\widehat{b}_{j}}\bigg(e_{i,j}-\lambda_{\widehat{b}}\widehat{b}_{j}\bigg);\\
	{u}_{i}&\leftarrow{u}_{i}+\gamma_u\bigg(e_{i,j}v_j-\lambda_u{u}_{i}\bigg);\\
	{v}_{j}&\leftarrow{v}_{j}+\gamma_v\bigg(e_{i,j}u_i-\lambda_v{v}_{j}\bigg);\\
	{w}_{j,j_{1}}&\leftarrow {w}_{j,j_{1}}+\gamma_w\left(\left|R^{K}(i;j)\right|^{-\frac{1}{2}}e_{i,j}(r_{i,j_{1}}-\overline{b}_{i,j_{1}})-\lambda_{w}{w}_{j,j_{1}}\right);\\
	{c}_{j,j_{2}}&\leftarrow {c}_{j,j_{2}}+\gamma_c\left(\left|N^{K}(i;j)\right|^{-\frac{1}{2}}e_{i,j}-\lambda_{c}{c}_{j,j_{2}}\right),
	\end{aligned}
	\right.\\
	\end{aligned}
	\label{sgd_update}
	\end{equation}
\end{small}
where the parameters
$\{ \gamma_{b_{i}}, \gamma_{\widehat{b}_{j}}, \gamma_{u}$, $\gamma_{v}, \gamma_w, \gamma_c \}$ are the corresponding learning rates and $e_{i,j}=r_{i,j}-\widehat{r}_{i,j}$.
The update rule (\ref{sgd_update}) has parallel inherence. Then,
the proposed CULSH-MF is comprised of the following three steps:

1) \textbf{Basic Optimization Structure} (\textbf{CUSGD++}):
CUSGD++ only considers the basic two parameters $\{\textbf{U}, \textbf{V}\}$.
Compared with cuSGD,
CUSGD++ has the following two advantages:
(1) Due to the higher usage of GPU registers in Stream Multiprocessors (SMs),
$u_i$ or $v_j$ can be updated in the registers,
avoiding the time overhead caused by a large number of memory accesses.
The memory access model is illustrated in Fig. \ref{memory_access}.
SM $\{1,2\}$ update $\{u_{1}, u_{2}\}$ in the registers, respectively;
and $\bigl\{
\{v_{1}, v_{3}, v_{4}, v_{7},v_{8},v_{11},v_{13}\},
\{v_{1}, v_{4}, v_{6}, v_{7},v_{9},v_{10},\\v_{12}\}
\bigl\}$ are returned to global memory after each update step.
(2) Due to the disentanglement of the parameters in the update rule (\ref{sgd_update}),
the data access conflict is reduced, which ensures a high access speed.
From the update rule (\ref{sgd_update}), the update processes of $\{\textbf{U}, \textbf{V}\}$ are symmetric.
Algorithm \ref{cumf-sgd++} only describes the update process of $\{\textbf{U}\}$ in the registers as follows:
\textbf{(1)} Lines $2-3$:
Given $TB$ thread blocks,
feature vectors $\bigl\{u_{i}|i\in \{1,\cdots,M\}\bigl\}$ are evenly assigned to thread blocks $\bigl\{TB_{tb\_idx}|tb\_idx\in\{1,\cdots,TB\}\bigl\}$.
Each thread block $TB_{tb\_idx}$ reads its own feature vector $u_i$ from the global memory into the registers.
\textbf{(2)} Line $4$:
The feature vector $u_i$ with all nonzero values $\{r_{i,j}| j\in {\Omega}_{i}\}$ in the thread block $TB_{tb\_idx}$ is updated.
\textbf{(3)} Lines $5-7$:
Use the \emph{warp} \emph{shuffle} \emph{instructions} \cite{ex117} to accelerate the dot product $u_{i}v_{j}^{T}$ of two vectors $\{u_{i}, v_{j}\}$ and broadcast the result.
This technology with additional hardware support uses registers that are faster than shared memory and does not involve thread synchronization.
Furthermore, this technology aligns and merges memory to reduce the access time.
The number of threads in a $\emph{thread}$ $\emph{warp}$ under the CUDA platform is 32, and
elements $\bigl\{u_{i,f},v_{j,f}|f\in \{1,\cdots,F\}\bigl\}$ in feature vectors $\{u_{i}, v_{j}\}$ are evenly assigned to thread blocks $\bigl\{T_{t\_idx}|t\_idx\in\{1,\cdots,32\}\bigl\}$.
A thread $T_{t\_idx}$ in each thread block $TB_{tb\_idx}$ sequentially reads the corresponding elements $\bigl\{u_{i,f},v_{j,f}|f\%32=t\_idx,f\in \{1,\cdots,F\}\bigl\}$ in feature vectors $\{u_{i}, v_{j}\}$, and the thread $T_{t\_idx}$
calculates the corresponding products $\bigl\{u_{i,f}v_{j,f}|f\%32=t\_idx,f\in \{1,\cdots,F\}\bigl\}$.
Then, the \emph{warp} \emph{shuffle} in the thread $T_{t\_idx}$ to obtain the dot product $u_{i}v_{j}^{T}=\sum\limits_{t\_idx}\sum\limits_{f\%32=t\_idx}u_{i,f}v_{j,f}$.
\textbf{(4)} Lines $8-10$:
Feature vectors $u_{i}$ are updated in the registers to avoid rereading from global memory for the next update,
and feature vectors $v_{j}$ are updated directly in global memory.
\textbf{(5)} Line $11$:
After all nonzero values $\{r_{i,j}| j\in {\Omega}_{i}\}$ have been updated,
the latest $u_{i}$ are written to global memory because it will no longer be used.

\begin{algorithm}[htbp]
	\caption{CUSGD++}
	\label{cumf-sgd++}
	\vspace{.1cm}
	$\mathcal{G}\{parameter\}$: parameter in global memory\\
	$\mathcal{R}\{parameter\}$: parameter in register memory\\
	$\textbf{Input}$:
	Initialization of
	low-rank feature matrices $\{\textbf{U}, \textbf{V}\}$,
	interaction matrix $\textbf{R}$,
	learning rate $\{\gamma_u, \gamma_v\}$,
	regularization parameter $\{\lambda_{u}, \lambda_{v}\}$, and
	training epoches $epo$.\\
	$\textbf{Output}$: $\textbf{U}$.\\
	\begin{algorithmic}[1]
		\FOR{: $loop$ from $1$ to $epo$}
		\FOR{$\textbf{(parallel)}$: $\bigl\{TB_{tb\_idx}|tb\_idx\in\{1,\cdots,TB\}\bigl\}$ manages its own feature vectors $\bigl\{u_{i}|i\in \{1,\cdots,M\}\bigl\}$}
		\STATE $\mathcal{R}\{u_{i}\}\leftarrow\mathcal{G}\{u_{i}\}$\\
		\FOR{: all $\bigl\{r_{i,j}|j \in {\Omega}_{i}\bigl\}$}
		\STATE Calculate $\widehat{r}_{i,j}=u_iv_j^{T}$.\\
		\STATE Calculate ${e}_{i,j}={r}_{i,j}-\widehat{r}_{i,j}$.\\
		\STATE Update $u_{i}, v_{j}$ by update rule (\ref{sgd_update}).\\
		\STATE $\mathcal{R}\{u_{i}\}\leftarrow u_{i}$\\
		\STATE $\mathcal{G}\{v_{j}\}\leftarrow v_{j}$\\
		\ENDFOR
		\STATE $\mathcal{G}\{u_{i}\}\leftarrow\mathcal{R}\{u_{i}\}$\\
		\ENDFOR
		\ENDFOR
	\end{algorithmic}
	
\end{algorithm}

2) \textbf{Aggregated Model} (\textbf{CULSH-MF}):
The updating process of $\{\textbf{W}, \textbf{C}\}$ for each thread $T_{t\_idx}$ is imbalanced.
This imbalance does not affect the serial model.
However, it obviously affects the running speed of the parallel model.
The most significant impacts are the following two points:
(1) discontinuous memory access, and
(2) imbalanced load on each thread $T$.
In order to solve the above problems,
an adjustment for the parameters $\{\textbf{W}, \textbf{C}\}$ is proposed in this section.
In CULSH-MF,
the adjustment takes the set $R(i)$ as a complement of the set $N(i)$.
Therefore, $S^{K}(j)=R^{K}(i;j)\bigcup N^{K}(i;j)$,$R^{K}(i;j)\bigcap N^{K}(i;j)=\varnothing$.
Thus,
the number of the involved elements for $\{\textbf{W}, \textbf{C}\}$ are equal and
each variable $J_{j}$ involves $2K$ parameters $\bigl\{\{w_{j,k}|k\in\{1,\cdots,K\}\},\{c_{j,k}|k\in\{1,\cdots,K\}\}\bigl\}$.
For the convenience of the expression, we use $k_1$ and $k_2$ to represent the indexes of $j_1$ and $j_2$ in these $K$ parameters, respectively,
which means that ${w}_{j,j_{1}}$ and ${c}_{j,j_{2}}$ are represented as ${w}_{j,k_{1}}$ and ${c}_{j,k_{2}}$, respectively.
The computational process of $\sum\limits_{{J_{j_{1}}\in R^{K}(i;j)}}(r_{i,j_{1}}-\overline{b}_{i,j_{1}}){w}_{j,k_{1}}$ and
$\sum\limits_{J_{j_{2}}\in N^{K}(i;j)}{c}_{j,k_{2}}$ involves the dot product and summation operations.
Thus, the $\emph{warp}$ $\emph{shuffle}$ $\emph{instructions}$, which can align and merge memory to reduce the overhead for GPU memory access, are used.

CULSH-MF also takes advantage of the register to reduce the memory access overhead and then increase the overall speed.
Due to the limited space, we only introduce the update rule of $\{\textbf{V}, \widehat{b}_{j}, \textbf{W}, \textbf{C}\}$ in the registers.
In Algorithm \ref{cumf-lsh}, the update process is presented in detail as follows:
\textbf{(1)} Line $1$:
Average value $\mu$ $=$ $\sum\limits_{(i,j)\in\Omega} r_{i,j}\bigl/|\Omega|$ as the basis value.
\textbf{(2)} Lines $3-7$:
Given TB thread blocks,
parameters $\{v_{j}, \widehat{b}_{j}, w_{j}, c_{j} |j\in \{1,\cdots,N\}\}$ are evenly assigned to thread blocks $\bigl\{TB_{tb\_idx}|tb\_idx\in\{1,\cdots,TB\}\bigl\}$.
Each thread block $TB_{tb\_idx}$ reads its own parameters $\{v_{j}, \widehat{b}_{j}, w_{j}, c_{j}\}$ from the global memory into the registers.
In addition, the reading of memory is also aligned and merged.
\textbf{(3)} Lines $8$:
The parameters $\{v_{j}, \widehat{b}_{j}, w_{j}, c_{j}\}$ with all nonzero values $\{r_{i,j}| i\in \widehat{\Omega}_{j}\}$ in thread block $TB_{tb\_idx}$ are updated.
\textbf{(3)} Lines $9-11$:
Use the \emph{warp} \emph{shuffle} \emph{instructions} \cite{ex117} to accelerate the dot product $u_{i}v_{j}^{T}$ and summation $\{\sum\limits_{j_{1}\in R^{K}(i;j)}(r_{i,j_{1}}-b_{i,j_{1}}){w}_{j,k_{1}},
\sum\limits_{j_{2}\in N^{K}(i;j)}{c}_{k,k_{2}}\}$.
Elements $\bigl\{u_{i,f},v_{j,f}, w_{j,k_1}, c_{j,k_2}
|f\in \{1,\cdots,F\}, k_1, k_2 \in \{1,\cdots,K\}\bigl\}$ in parameters $\{u_{i}, v_{j}, w_{j}, c_{j} |j\in \{1,\cdots,N\}\}$ are evenly assigned to thread blocks $\bigl\{T_{t\_idx}|t\_idx\in\{1,\cdots,32\}\bigl\}$.
A thread $T_{t\_idx}$ in each thread block $TB_{tb\_idx}$ sequentially reads the corresponding elements $\bigl\{u_{i,f},v_{j,f}, w_{j,k_1},$ $c_{j,k_2}|f\%32=k_1\%32=k_2\%32=t\_idx,f\in \{1,\cdots,F\}, k_1, k_2 \in \{1,\cdots,K\}\bigl\}$
in parameters $\{u_{i}, v_{j}, w_{j}, c_{j}\}$, and the thread $T_{t\_idx}$
calculates the corresponding calculations $\bigl\{u_{i,f}v_{j,f},(r_{i,j_{1}}-b_{i,j_{1}}){w}_{j,k_{1}},{c}_{k,k_{2}}
|f\%32=k_1\%32=k_2\%32=t\_idx,f\in \{1,\cdots,F\}, k_1, k_2 \in \{1,\cdots,K\}\bigl\}$.
Please note that since $S^{K}(j)=R^{K}(i;j)$ $\bigcup N^{K}(i;j)$ and $R^{K}(i;j)\bigcap N^{K}(i;j)=\varnothing$, the thread only calculates one of $(r_{i,j_{1}}-b_{i,j_{1}}){w}_{j,k_{1}}$ and ${c}_{k,k_{2}}$.
This makes the load of each thread $T_{t\_idx}$ relatively balanced during the update process.
Then, the \emph{warp} \emph{shuffle} in thread $T_{t\_idx}$ to obtain the
$\widehat{r}_{i,j}$ $=$ $\mu + b_{i} + \widehat{b}_{j}$ $+\sum\limits_{t\_idx}\bigl(\sum\limits_{f\%32=t\_idx}u_{i,f}v_{j,f} + \sum\limits_{k_1\%32=t\_idx\atop j_{1}\in R^{K}(i;j)}(r_{i,j_{1}}-b_{i,j_{1}}){w}_{j,k_{1}} + \sum\limits_{k_2\%32=t\_idx\atop j_{2}\in N^{K}(i;j)}{c}_{k,k_{2}}\bigl)$.
\textbf{(4)} Lines $12-18$:
Parameters $\{v_{j}, \widehat{b}_{j}, w_{j}, c_{j}\}$ are updated in the registers to avoid rereading from global memory for the next update,
and parameters $\{u_{i}, {b}_{i}\}$ are updated directly in global memory.
\textbf{(5)} Lines $19-22$:
After all nonzero values $\{r_{i,j}| i\in \widehat{{\Omega}}_{j}\}$ have been updated,
the latest $\{v_{j}, \widehat{b}_{j}, w_{j}, c_{j}\}$ are written to global memory because they will no longer be used.
These operations are similar to CUSGD++.

The algorithm has the following advantages:
\textbf{(1)} It stores a large number of parameters in registers,
avoiding frequent access to global memory and decreasing the time consumption; and
\textbf{(2)} The parameter distribution is regular
such that each thread $T_{t\_idx}$ is balanced,
which can avoid idle threads and can improve the active rate of threads.
Compared with CUSGD++, CULSH-MF can assemble more tangled parameters of the nonlinear MF model.
The parameters $\{v_{j}, \widehat{b}_{j}, w_{j}, c_{j}\}$ are taken as a whole, and the memory is merged and aligned.
Then, the use of $\emph{warp}$ $\emph{shuffle}$ can further optimize the memory access by allowing the computational overhead to be further reduced.
The spatial overhead is $O(|\Omega|+MF+NF+3NK)$ for interaction sparse matrix $\textbf{R}$, low-rank factor matrices $\{\textbf{U}, \textbf{V}\}$, influence matrices $\{\textbf{W}, \textbf{C}\}$ and the Top-$K$ GSM matrix $\textbf{J}^{K}$.
\begin{algorithm}[htbp]
	\caption{CULSH-MF}
	\label{cumf-lsh}
	\vspace{.1cm}
	$\mathcal{G}\{parameter\}$: parameter in global memory\\
	$\mathcal{R}\{parameter\}$: parameter in register memory\\
	$\textbf{Input}$:
	Initialization for $\{\textbf{U}, \textbf{V}, \mu, b_{i}, \widehat{b}_{j}, \textbf{W}, \textbf{C}\}$,
	sparse matrix $\textbf{R}$,
	learning rate parameters $\{\gamma_{b}, \gamma_{\widehat{b}}, \gamma_{u}$, $\gamma_{v}, \gamma_w, \gamma_c \}$,
	regularization parameters $\{\lambda_{b}, \lambda_{\widehat{b}}, \lambda_{u}, \lambda_{v}, \lambda_{w}, \lambda_{c}\}$, and training epoches $epo$.\\
	$\textbf{Output}$: $\{\textbf{U}, \textbf{V}, \mu, b_{i}, \widehat{b}_{j}, \textbf{W}, \textbf{C}\}$.\\
	\begin{algorithmic}[1]
		\STATE $u$ $\leftarrow$ Average value of rating matrix $\textbf{R}$.\\
		\FOR{$loop$ from $1$ to $epo$}
		\FOR{$\textbf{(parallel)}$: $\bigl\{TB_{tb\_idx}|tb\_idx\in\{1,\cdots,TB\}\bigl\}$ manages its own parameters $\{u_{i}, v_{j}, w_{j}, c_{j} |j\in \{1,\cdots,N\}\}$}
		\STATE $\mathcal{R}\{\widehat{b}_{j}\}\leftarrow\mathcal{G}\{\widehat{b}_{j}\}$;\\
		\STATE $\mathcal{R}\{v_{i}\}\leftarrow\mathcal{G}\{v_{j}\}$\\
		\STATE $\mathcal{R}\{w_{j}\}\leftarrow\mathcal{G}\{w_{j}\}$\\
		\STATE $\mathcal{R}\{c_{j}\}\leftarrow\mathcal{G}\{c_{j}\}$\\
		\FOR{all $\{r_{i,j}|i \in \widehat{\Omega}_{j}\}$}
		\STATE Calculate $\widehat{r}_{i,j}$ by equation (\ref{rr}).\\
		\STATE Calculate ${e}_{i,j}={r}_{i,j}-\widehat{r}_{i,j}$.\\
		\STATE Update $\{b_{i}, \widehat{b}_{j}, u_{i}, v_{j}, w_{j}, c_{j}\}$ by update rule (\ref{sgd_update}).\\
		\STATE $\mathcal{R}\{\widehat{b}_{j}\}\leftarrow \widehat{b}_{j}$\\
		\STATE $\mathcal{R}\{v_{j}\}\leftarrow v_{j}$\\
		\STATE $\mathcal{R}\{w_{j}\}\leftarrow w_{j}$\\
		\STATE $\mathcal{R}\{c_{j}\}\leftarrow c_{j}$\\
		\STATE $\mathcal{G}\{b_{i}\}\leftarrow b_{i}$\\
		\STATE $\mathcal{G}\{u_{i}\}\leftarrow u_{i}$\\
		\ENDFOR
		\STATE $\mathcal{G}\{\widehat{b}_{j}\}\leftarrow\mathcal{R}\{\widehat{b}_{j}\}$\\
		\STATE $\mathcal{G}\{v_{j}\}\leftarrow\mathcal{R}\{v_{j}\}$\\
		\STATE $\mathcal{G}\{w_{j}\}\leftarrow\mathcal{R}\{w_{j}\}$\\
		\STATE $\mathcal{G}\{c_{j}\}\leftarrow\mathcal{R}\{c_{j}\}$\\		
		\ENDFOR	
		\ENDFOR
	\end{algorithmic}
\end{algorithm}

\begin{figure}[htbp]
	\centering
	\includegraphics[width=6.0in]{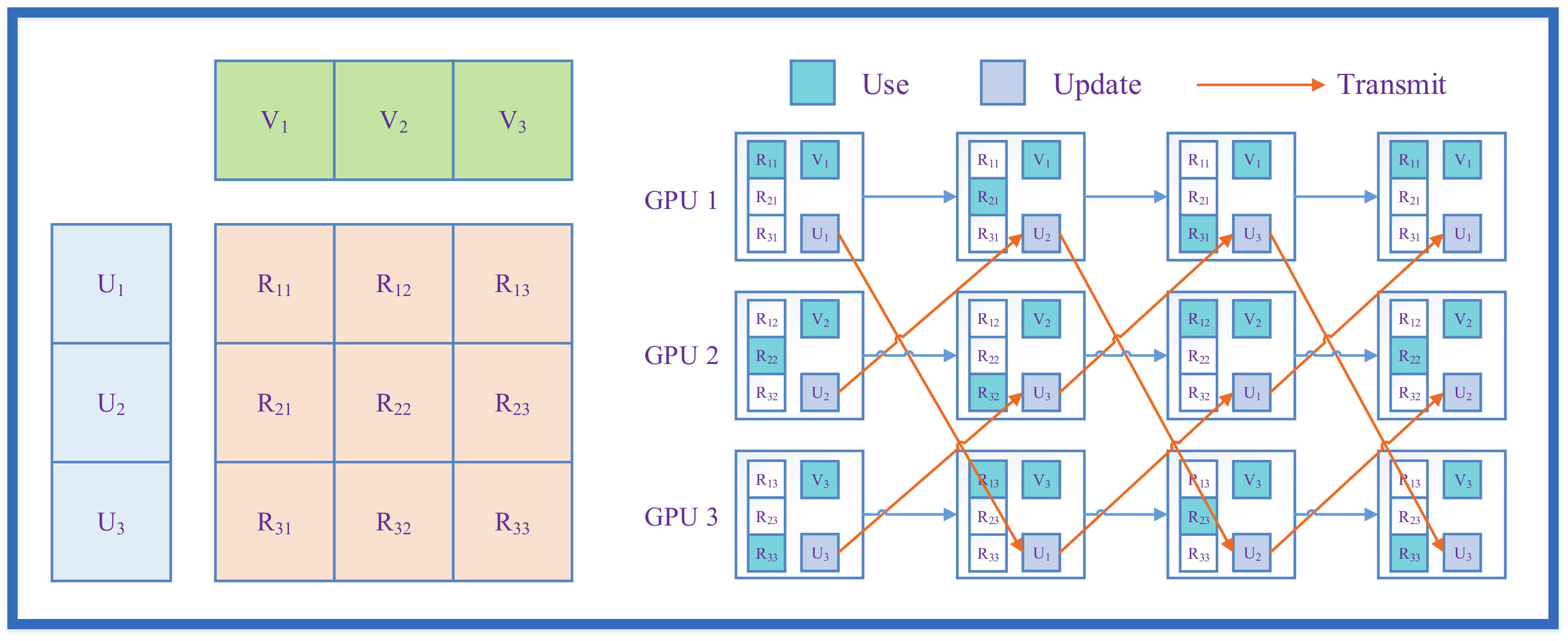}
	\caption{Multi-GPU Solution}
	\label{fig_multi_gpu}
\end{figure}

3) \textbf{Multi-GPU Model}: With big data,
a single GPU still cannot meet our requirements.
Therefore, the method must be extended to multiple GPUs (MCUSGD++/MCULSH-MF).
We use data parallelism to allow multiple GPUs to run our algorithms at the same time.
To avoid data conflicts,
each GPU-updated block cannot be on the same $I_i$ or on the same $J_j$.
After the update is completed,
the updated parameters are not sent back to the CPU because another GPU needs these data directly.
Transferring data directly in the GPUs avoids the extra time overhead of uploading to the CPU and then allocates them to other GPUs.
Each GPU is assigned some specific parameters, which are not needed by other GPUs.
After all updates are completed,
each GPU passes the parameters that are stored at that time back to the CPU.

Assume that we have $D$ GPUs.
The sparse matrix $\textbf{R}$ is divided into
$D\times D$ parts $\bigl\{\textbf{R}_{d_1,d_2}|d_1,d_2\in\{1,\cdots,D\}\bigl\}$.
Low-rank feature matrices $\{\textbf{U}, \textbf{V}\}$ are divided into $\bigl\{ \{\textbf{U}_{d_1}|d_1\in \{1,\cdots,D\}\}, \{\textbf{V}_{d_2}|d_2\in \{1,\cdots,D\}\} \bigl\}$, respectively.
Influence matrices $\{\textbf{W}, \textbf{C}\}$ are divided into $\bigl\{ \{\textbf{W}_{d_2}|d_2\in \{1,\cdots,D\}\}, \{\textbf{C}_{d_2}|d_2\in \{1,\cdots,D\}\} \bigl\}$, respectively.
The parameters $\bigl\{\textbf{R}_{d_1,d_2}, \textbf{V}_{d_2}, \textbf{W}_{d_2}, \textbf{C}_{d_2}|d_1\in \{1,\cdots,D\}\bigl\}$ are allocated to the $d_2$th GPU and do not require transmission.
Parameter $\textbf{V}_{d_1}$ is allocated to the $d_1$th GPU at initialization and then transferred to another GPU after each update step.
Fig. \ref{fig_multi_gpu} depicts MCUSGD++ on three GPUs.
MCULSH-MF is similar and is given in parentheses below.
The sparse matrix $\textbf{R}$ is divided into $3\times3$ blocks.
The training process of all the parameters is divided into three parts:
(1): GPUs $\{1,2,3\}$ update
$\bigl\{
\{\textbf{U}_{1}, \textbf{V}_{1}, (\textbf{W}_{1}, \textbf{C}_{1})\},
\{\textbf{U}_{2}, \textbf{V}_{2}, (\textbf{W}_{2}, \textbf{C}_{2})\},
\{\textbf{U}_{3}, \textbf{V}_{3}, (\textbf{W}_{3}, \textbf{C}_{3})\}
\bigl\}$
and then transmit
$\bigl\{
\textbf{U}_{1},
\textbf{U}_{2},
\textbf{U}_{3}
\bigl\}$
to GPUs $\{3, 1, 2\}$, respectively;
(2): GPUs $\{1,2,3\}$ update
$\bigl\{
\{\textbf{U}_{2}, \textbf{V}_{1}, (\textbf{W}_{1}, \textbf{C}_{1})\},
\{\textbf{U}_{3}, \textbf{V}_{2}, (\textbf{W}_{2}, \textbf{C}_{2})\},
\{\textbf{U}_{1}, \textbf{V}_{3}, (\textbf{W}_{3},\\ \textbf{C}_{3})\}
\bigl\}$
and transmit
$\bigl\{
\textbf{U}_{2},
\textbf{U}_{3},
\textbf{U}_{1}
\bigl\}$ to GPUs $\{3, 1, 2\}$, respectively; and
(3): GPUs $\{1,2,3\}$ update
$\bigl\{
\{\textbf{U}_{3}, \textbf{V}_{1},\\ (\textbf{W}_{1}, \textbf{C}_{1})\},
\{\textbf{U}_{1}, \textbf{V}_{2}, (\textbf{W}_{2}, \textbf{C}_{2})\},
\{\textbf{U}_{2}, \textbf{V}_{3}, (\textbf{W}_{3}, \textbf{C}_{3})\}
\bigl\}$
and transmit
$\bigl\{
\textbf{U}_{3},
\textbf{U}_{1},
\textbf{U}_{2}
\bigl\}$ to GPUs $\{3, 1, 2\}$, respectively.

\subsection{Online Learning}
Big data analysis should consider the incremental data, and the
corresponding model can be compatible with the incremental data.
The amount of incremental data is much smaller than the amount of original data.
Thus, the time overhead for retraining the overall data is not worthwhile.
It is nontrivial to design an online model for incremental data.
The variable sets $\{I, \overline{I}, \widehat{I} \}$ and $\{J, \overline{J}, \widehat{J}\}$ are denoted as the original variable set,
new variable set,
and overall variable set, respectively.
In this work,
we consider that the new variable sets $\overline{I}$ and $\overline{J}$ enter the system and interact with variable sets $J$ and $I$, respectively.
Please note that this allows variable set $\overline{I}$ to interact with variable set $\overline{J}$.

For the original variable $J_j$ $\in$ $J$,
the Top-$K$ nearest neighbours $\{J_{j_{1}}, \cdots, J_{j_{K}}\}$
$\in$ $J$ are kept.
For the new variable $\overline{J}_{\overline{j}}$ $\in$ $\overline{J}$, we search its Top-$K$ nearest neighbours $\{\widehat{J}_{\widehat{j}_{1}}, \cdots, \widehat{J}_{\widehat{j}_{K}}\}$ $\in$ $\widehat{J}$.
The hash value of variable set $J$ depends on $I$,
and the hash value of variable set $\overline{J}$ depends on $\widehat{I}$.
In order to keep them consistent,
we update the hash value of variable $J_j$ $\in$ $J$; then,
we save the intermediate variables $\sum_{i\in \widehat{\Omega}_{j}}\Psi(r_{i,j}) (2 \cdot H_{i}-1)$ of simLSH
and update $\overline{H}_{j}=\Upsilon\bigg(
	\sum_{i\in \widehat{\Omega}_{j}}\Psi(r_{i,j})\Phi (H_{i})+
	\sum_{\overline{i}\in \widehat{\Omega}_{j}}\Psi(r_{\overline{i},j})\Phi(H_{\overline{i}})\bigg)$.
Furthermore, we obtain
$\overline{H}_{\overline{j}}=\Upsilon$  $\bigl($
	$\sum_{\widehat{i}\in \widehat{\Omega}_{\overline{j}}}\Psi(r_{\widehat{i},\overline{j}})$ $\Phi(H_{\widehat{i}})$ $\bigl)$.
The online learning solution is described in Algorithm \ref{online} as follows:
(1) Lines $1-3$:
Update the hash value $\overline{H}_{j}$ for variable $J_{j}$ $\in$ $J$.
Saving the intermediate variables makes the process only require a small amount of calculation.
(2) Lines $4-6$:
Calculate hash value $\overline{H}_{\overline{j}}$ for variable $\overline{J}_{\overline{j}}$ $\in$ $\overline{J}$.
Both the hash value of variable set $J$ and the hash value of variable set $\overline{J}$ depend on $\widehat{I}$.
(3) Lines $7-9$:
Search the Top-$K$ nearest neighbours $\{\widehat{J}_{\widehat{j}_{1}}, \cdots, \widehat{J}_{\widehat{j}_{K}}\}$ of variable $\overline{J}_{\overline{j}}$ $\in$ $\overline{J}$.
The Top-$K$ nearest neighbours in the overall variable set $\widehat{J}$ can provide more information.
(4) Lines $10-12$: Update $\{b_{\overline{i}}, u_{\overline{i}}\}$ for variable $\overline{I}_{\overline{i}}$ $\in$ $\overline{I}$.
$\{r_{\overline{i},j}|\overline{I}_{\overline{i}}\in \overline{I}, J_j \in J\}$
is used and $\{\widehat{b}_{j}, v_{j}, w_{j}, c_{j}\}$ remains unchanged, but they can still be stored in registers to reduce memory access.
(5) Lines $13-15$: Updating $\{\widehat{b}_{\overline{j}}, v_{\overline{j}}, w_{\overline{j}}, c_{\overline{j}}\}$ for variable $\overline{J}_{\overline{j}}$ $\in$ $\overline{J}$, $\{r_{\widehat{i},\overline{j}}|\widehat{I}_{\widehat{i}}\in \widehat{I}, \overline{J}_{\overline{j}} \in \overline{J}\}$
is used, and $\{\widehat{b}_{j}, v_{j}, w_{j}, c_{j}\}$ remains unchanged.

\begin{algorithm}[htbp]
	\caption{Online Learning}
	\label{online}
	\vspace{.1cm}
	$\textbf{Input}$: $\{b_{i}, u_{i}, \widehat{b}_{j}, v_{j}, w_{j}, c_{j}\}$, new variable sets $\overline{I}$ and $\overline{J}$, random Hash values $H_{\overline{i}}$.\\
	$\textbf{Output}$: $\{b_{\overline{i}}, u_{\overline{i}}\}$,$\{\widehat{b}_{\overline{j}}, v_{\overline{j}}, w_{\overline{j}}, c_{\overline{j}}\}$.\\
	
	\begin{algorithmic}[1]
		\FOR{$loop$ from $1$ to $epo$}
		\FOR{$\textbf{(parallel)}$: Variables $J_j$ $\in$ $J$ are evenly assigned to thread blocks $\bigl\{TB_{tb\_idx}|tb\_idx\in\{1,\cdots,TB\}\bigl\}$}
		\STATE Update the hash value $\overline{H}_{j}$ for variable $J_{j}$ $\in$ $J$.
		\ENDFOR
		\FOR{$\textbf{(parallel)}$: Variables $\overline{J}_{\overline{j}}$ $\in$ $\overline{J}$ are evenly assigned to thread blocks $\bigl\{TB_{tb\_idx}|tb\_idx\in\{1,\cdots,TB\}\bigl\}$}
		\STATE 	Calculate the hash value $\overline{H}_{\overline{j}}$ for variable $\overline{J}_{\overline{j}}$ $\in$ $\overline{J}$.
		\ENDFOR
		\FOR{$\textbf{(parallel)}$: Variables $\overline{J}_{\overline{j}}$ $\in$ $\overline{J}$ are evenly assigned to thread blocks $\bigl\{TB_{tb\_idx}|tb\_idx\in\{1,\cdots,TB\}\bigl\}$}
		\STATE 	Search the Top-$K$ nearest neighbours $\{\widehat{J}_{\widehat{j}_{1}}, \cdots, \widehat{J}_{\widehat{j}_{K}}\}$ of the variable $\overline{J}_{\overline{j}}$ $\in$ $\overline{J}$.
		\ENDFOR
		\FOR{$\textbf{(parallel)}$: Variables $J_j$ $\in$ $J$ are evenly assigned to thread blocks $\bigl\{TB_{tb\_idx}|tb\_idx\in\{1,\cdots,TB\}\bigl\}$}
		\STATE 	Update $\{b_{\overline{i}}, u_{\overline{i}}\}$ for variable $\overline{I}_{\overline{i}}$ $\in$ $\overline{I}$.
		\ENDFOR
		\FOR{$\textbf{(parallel)}$: Variables $\overline{J}_{\overline{j}}$ $\in$ $\overline{J}$ are evenly assigned to thread blocks $\bigl\{TB_{tb\_idx}|tb\_idx\in\{1,\cdots,TB\}\bigl\}$}
		\STATE 	Update $\{\widehat{b}_{\overline{j}}, v_{\overline{j}}, w_{\overline{j}}, c_{\overline{j}}\}$ for variable $\overline{J}_{\overline{j}}$ $\in$ $\overline{J}$.
		\ENDFOR
		\ENDFOR
	\end{algorithmic}
	
\end{algorithm}

\section{Experiments}
CULSH-MF is comprised of two parts:
1) Basic parallel optimization model depends on CUSGD++,
which can utilize the GPU registers more and
disentangle the involved parameters.
CUSGD++ achieves the fastest speed compared to the state-of-the-art algorithms.
2) The Top-$K$ nearest neighbourhood query relies on the proposed simLSH, which can reduce the time and memory overheads.
Furthermore, it can improve the overall approximation accuracy.
In order to demonstrate the effectiveness of the proposed model,
we present the experimental settings in Section 5.1.
The speedup performance of CUSGD++ compared with the state-of-the-art algorithms is shown in Section 5.2.
The accuracy, robustness, online learning and multiple GPUs of CULSH-MF are presented in Section 5.3.
CULSH-MF is a nonlinear neighbourhood model for low-rank representation learning, and
we compare CULSH-MF with the DL model in Section 5.4 to demonstrate the effectiveness of CULSH-MF.

\subsection{Experimental Setting}
The experiments were run on an NVIDIA Tesla P100 GPU with CUDA version 10.0.
The same software and hardware conditions can better reflect the superiority of the proposed algorithm.
The experiments are conducted on 3 public datasets:
Netflix \footnotemark[1] \footnotetext[1]{https://www.netflixprize.com/}, MovieLens \footnotemark[2] \footnotetext[2]{https://grouplens.org/datasets/movielens/} and
Yahoo! Music \footnotemark[3] \footnotetext[3]{https://webscope.sandbox.yahoo.com/}.
For MovieLens and Yahoo! Music, data cleaning is conducted, and $0$ values are changed from $0$ to $0.5$.
This will make cuALS work properly, which is one of the shortcomings of cuALS.
The specific situations of the datasets are shown in Table \ref{Data sets}.
The ratings in the Yahoo! Music dataset are relatively large,
which affects the training process.
In the actual training process, we divided all the ratings in the Yahoo! Music dataset by 20,
and then we multiply by 20 when verifying the results.
In this way, the ratings of the three datasets are in the same interval,
which facilitates the parameter selection.
The accuracy is measured by the $RMSE$ as:
\begin{equation}
\begin{aligned}
\setlength{\abovedisplayskip}{0pt}
\setlength{\belowdisplayskip}{0pt}
RMSE&=\sqrt{\bigg(\sum_{(i,j)\in\Gamma}(v_{i,j}-\widetilde{v}_{i,j})^{2}\bigg)\bigg/|\Gamma|},
\end{aligned}
\end{equation}
where $\Gamma$ denotes the test sets.

\begin{table}[htbp]
	\centering
	\footnotesize
	\setlength{\abovecaptionskip}{0pt}
	\caption{Data sets}
	\begin{tabular}{c|ccc}
		\hline \hline
		Parameter & Netflix  & Movielens & Yahoo!Music \\
		\hline
		M           & 480, 189   & 69, 878     & 586, 250      \\
		N           & 17, 770    & 10, 677     & 12, 658       \\
		$|\Omega|$  & 99, 072, 112 & 9, 900, 054   & 91, 970, 212    \\
		$|\Gamma|$   & 1, 408, 395  & 100, 000    & 1, 000, 000     \\
		Max Value & 5        & 5         & 100         \\
		Min Value & 1        & 0.5       & 0.5         \\
		\hline
		\hline
	\end{tabular}
	\label{Data sets}
\end{table}

The number of threads in a $\emph{thread}$ $\emph{warp}$ under the CUDA system is 32.
Therefore, we set the number of threads in the thread block to a multiple of 32.
This is done to maximize the utilization of the warp.
Then, in order to align access, we set the parameters $\{F, K\}$ as multiples of 32.

\subsection{CUSGD++}

CUSGD++ is used to compare cuALS \cite{ex148} and cuSGD \cite{ex149} on the three datasets.
The parameters of cuALS and cuSGD were set as described in their papers and optimized according to the hardware environment, and CUSGD++ uses the dynamic learning rate in \cite{ex165} as
\begin{equation}
\begin{split}
\gamma_t=\frac{\alpha}{1+\beta \cdot t^{1.5}},
\end{split}
\label{learn_rate}
\end{equation}
where the parameters $\{\alpha, \beta, t, \gamma_t\}$ represent the initial learning rate, adjusting parameter of the learning rate, the number of current iterations, and the learning rate at $t$ iterations, respectively.
The learning rate and other parameters in CUSGD++ are listed in Table \ref{CUMF-SGD++ Parameters}.

\begin{table}[htbp]
	\centering
	\footnotesize
	\setlength{\abovecaptionskip}{0pt}
	\caption{CUSGD++ Parameters}
	\begin{tabular}{c|ccc}
		\hline \hline
		Parameter & Netflix & Movielens & Yahoo!Music \\
		\hline
		$\alpha$     & 0.04    & 0.04      & 0.01        \\
		$\beta$      & 0.3     & 0.3       & 0.1         \\
		$\lambda_u$  & 0.035   & 0.035     & 0.02        \\
		$\lambda_v$  & 0.035   & 0.035     & 0.02        \\
		\hline
		\hline
	\end{tabular}
	\label{CUMF-SGD++ Parameters}
\end{table}

\begin{figure*}[htbp]
	\centering
	\includegraphics[width=1.80in]{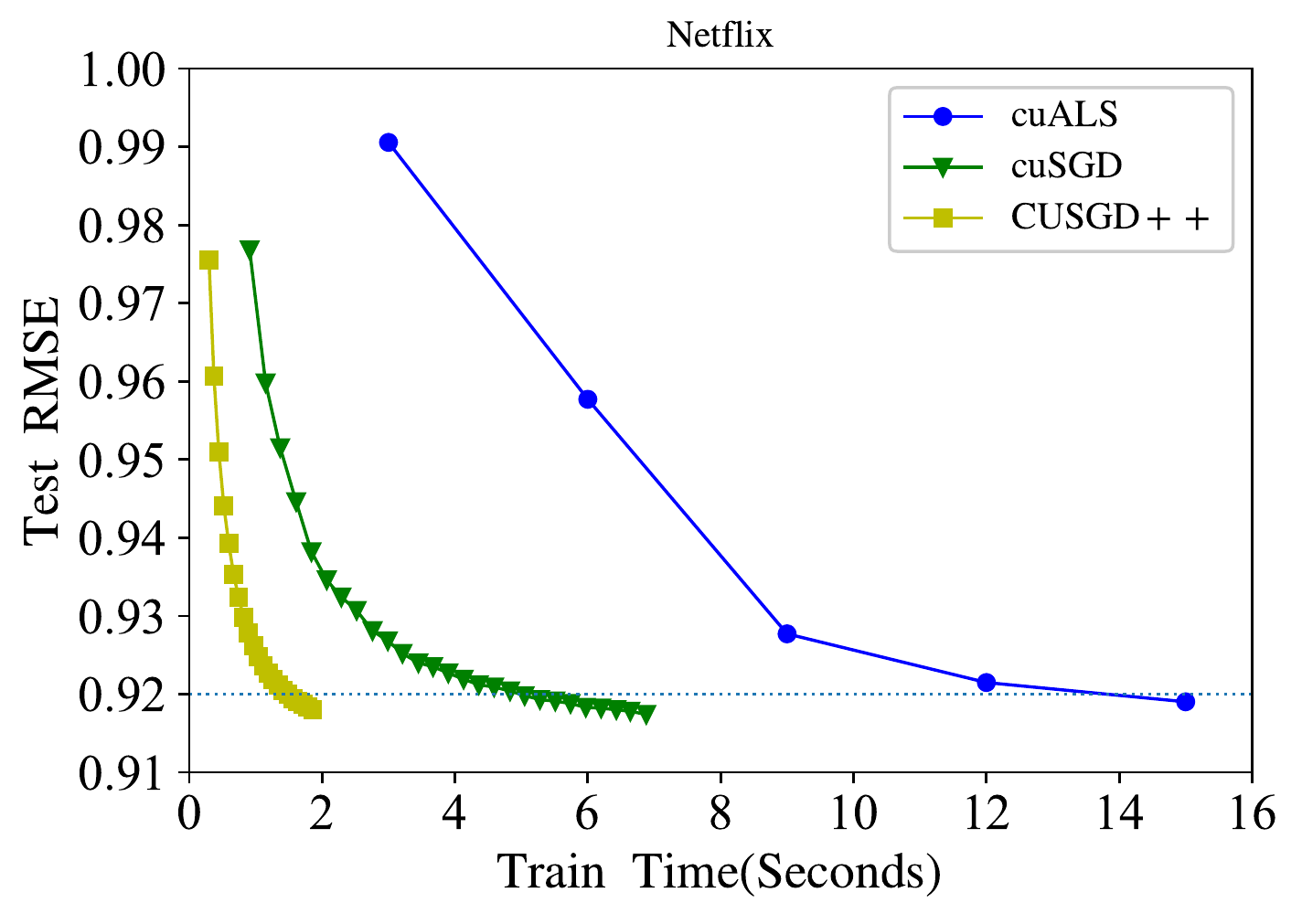}
	\includegraphics[width=1.80in]{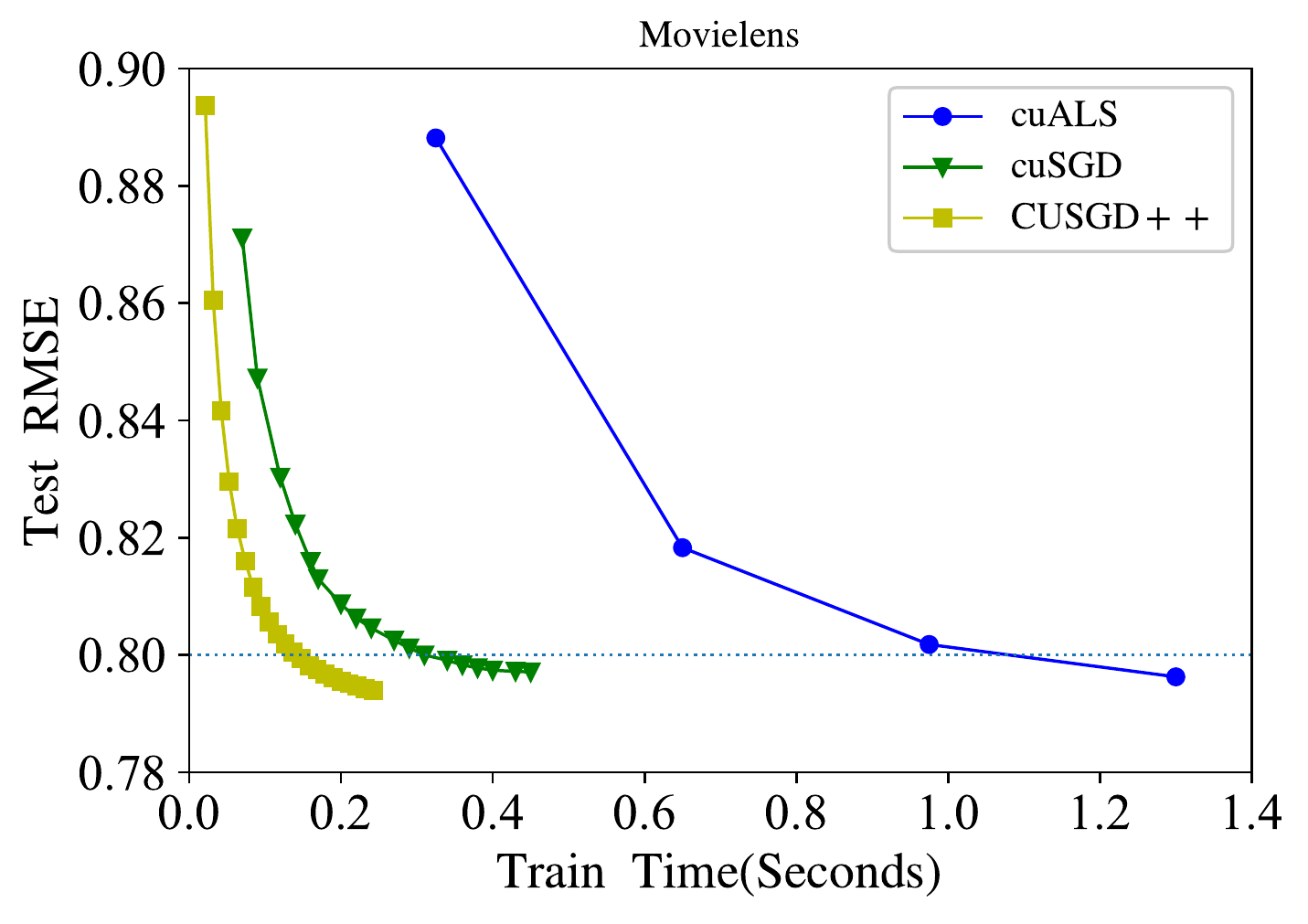}
	\includegraphics[width=1.80in]{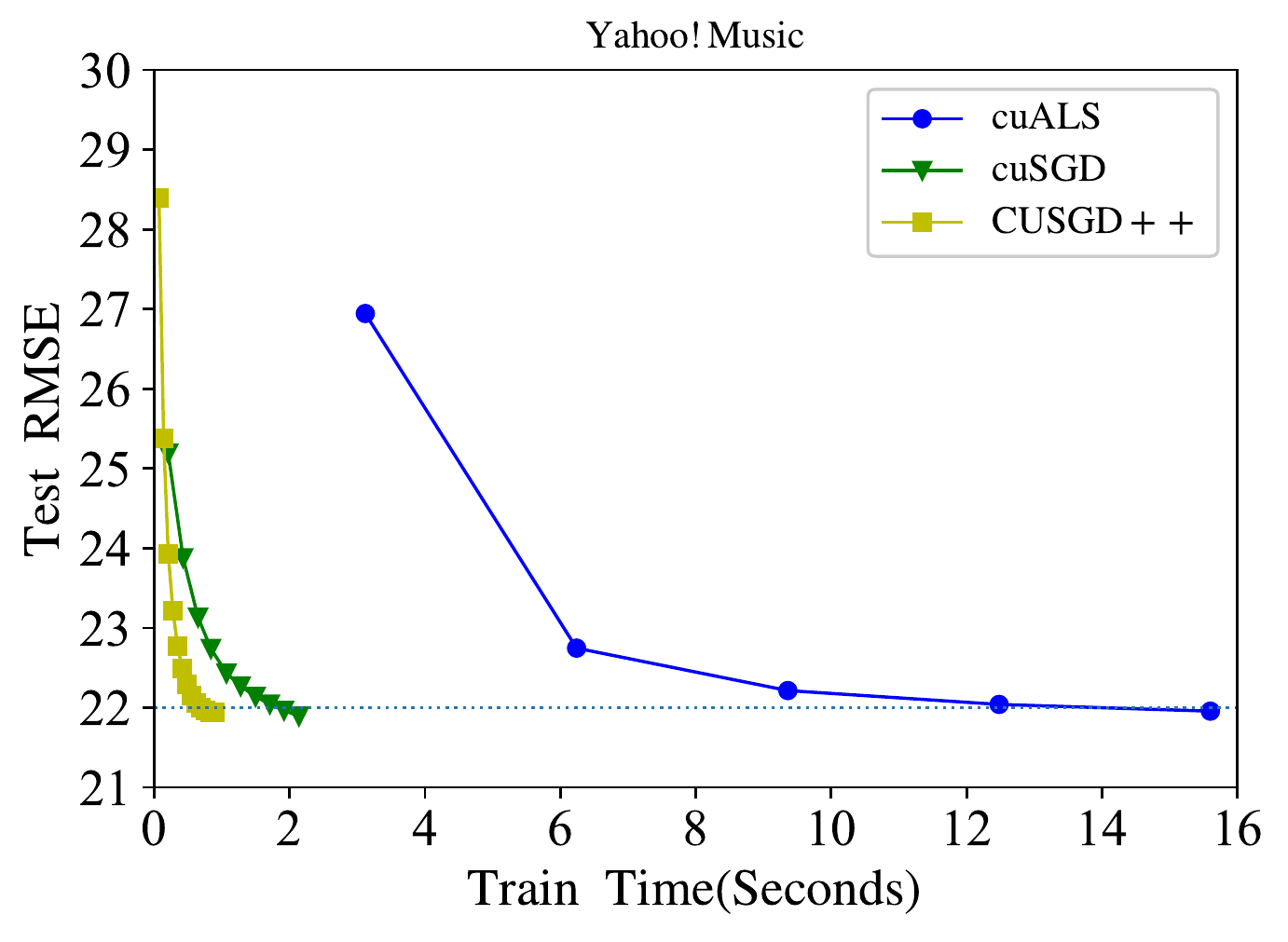}
	\caption{RMSE vs time: the experimental results demonstrate that CUSGD++ converges faster than other approaches.}
	\label{fig_cumd-dg++}
\end{figure*}

The GPU experiments are conducted on three datasets.
In order to ensure running fairness,
we ensure that the GPU executes these algorithms independently, and there is no other work.
Fig. \ref{fig_cumd-dg++} shows the relationship between the $RMSE$ and training time.
In Table \ref{Speed comparison based on cuALS}, the times it takes to achieve an acceptable $RMSE$ (0.92, 0.80, and 22.0 for Netflix, MovieLens and Yahoo! Music, respectively) are presented.
cuALS has an extremely fast descent speed, but the time of each iteration is very long
because the matrix inversion calculation is performed twice for each iteration.
Furthermore,
because the number of $\bigl\{r_{i,j}|j\in {\Omega}_{i}\bigl\}$ for each $I_i$ is very different
and the number of $\{r_{i,j}|i \in \widehat{\Omega}_{j}\}$ for each $J_j$ is the same,
the thread load imbalance further increases the time overhead.
cuSGD has a slower descent speed but less time overhead per iteration
due to using data parallelism without load balancing issues.

cuSGD has an obvious flaw in that it does not take full advantage of the hardware resources of the GPU.
cuSGD stores data in global memory, which makes it take too much time to read and write data.
Our proposed CUSGD++ is significantly faster than the state-of-the-art algorithms on the GPU.
CUSGD++ and cuSGD have the same number of iterations to obtain an acceptable $RMSE$, and
the speed of a single iteration is $2-3$ times faster than cuSGD.
With the same gradient descent algorithm,
the proposed CUSGD++ and cuSGD algorithms are basically the same in terms of descent speed.
CUSGD++ makes full use of the GPU hardware.
Therefore, the time overhead of each iteration is only approximately $1/3$ that of cuSGD.
It is inevitable that CUSGD++ results in a thread load imbalance problem, and our further work is to solve this problem.
Simultaneously, we simply sort the index of the row or column for $I_i$ $\in$ $I$ according to the number of $\bigl\{r_{i,j}|j\in {\Omega}_{i}\bigl\}$.
Therefore, $I_i$ containing more nonzero elements $\bigl\{r_{i,j}|j\in {\Omega}_{i}\bigl\}$ is updated first.
This approach can reduce the time overhead on a single iteration and achieve speedups of $\{1.02X, 1.03X, 1.06X\}$ on the Netflix, MovieLens and Yahoo! Music datasets, respectively.

\begin{table}[htbp]
	\centering
	\footnotesize
	\setlength{\abovecaptionskip}{0pt}
	\caption{Speedup comparison on the baseline cuALS}
	\begin{tabular}{c|ccc}
		\hline \hline
		Algorithm & Netflix     & Movielens   & Yahoo!Music \\
		\hline
		cuALS     & 15.00       & 1.30        & 15.60       \\
		cuSGD     & 5.05 (3.0X)  & 0.31 (4.2X)  & 1.92 (8.1X)  \\
		CUSGD++   & 1.49 (10.1X) & 0.15 (8.7X)  & 0.69 (22.6X) \\
		\hline \hline
	\end{tabular}
	\label{Speed comparison based on cuALS}
\end{table}

\subsection{CULSH-MF}

Before introducing the experiment, we will introduce the selection of the relevant parameters.
CULSH-MF still uses the dynamic learning rate in Equation (\ref{learn_rate}). The initial learning rate and regularization parameters are shown in Table \ref{The_CUMF_LSH}, and $\beta$ for all three datasets is $0.3$.
\begin{table}[htbp]
	\centering
	\footnotesize
	\setlength{\abovecaptionskip}{0pt}
	\caption{The initial learning speed and regularization parameters of CULSH-MF for all three datasets}
	\begin{tabular}{c|ccc}
		\hline
		\hline
		Parameter & Netflix & Movielens & Yahoo!Music \\ \hline
		$\alpha_{i}$  & 0.02    & 0.035     & 0.02        \\
		$\widehat{\alpha}_{j}$  & 0.02    & 0.035     & 0.02        \\
		$\alpha_u$           & 0.02    & 0.035     & 0.02        \\
		$\alpha_v$           & 0.02    & 0.035     & 0.02        \\
		$\alpha_w$           & 0.001   & 0.002     & 0.001       \\
		$\alpha_c$           & 0.001   & 0.002     & 0.001       \\
		$\lambda_{b_{i}}$    & 0.01    & 0.02      & 0.02        \\
		$\lambda_{\widehat{b}_{j}}$    & 0.01    & 0.02      & 0.02        \\
		$\lambda_u$          & 0.01    & 0.02      & 0.02        \\
		$\lambda_v$          & 0.01    & 0.02      & 0.02        \\
		$\lambda_w$          & 0.05    & 0.002     & 0.05        \\
		$\lambda_c$          & 0.05    & 0.002     & 0.05        \\ \hline \hline
	\end{tabular}
	\label{The_CUMF_LSH}
\end{table}
In order to clarify the superiority of CULSH-MF,
the experimental presentation is split into the following $5$ parts:
1) The overall performance comparison,
2) The performance comparison for the various methods of Top-$K$ nearest neighbourhood query,
3) The performance comparison of neighbourhood nonlinear MF with naive MF methods,
4) The performance comparison on a GPU and multiple GPUs, and
5) The robustness of CULSH-MF.

We first compare the serial algorithms, i.e., LSH-MF and GSM-based Top-$K$ nearest neighbourhood MF \cite{ex156}.
To ensure the fairness of the comparison, the parameters used are the same\cite{ex156}.
The serial algorithms are conducted on an Intel Xeon E5-2620 CPU, and
the CUDA parallelization algorithms are conducted on an NVIDIA Tesla P100 GPU.
Parameters $\{F, K\}$ are set as $\{32, 32\}$, respectively.
Table \ref{CUMF-LSH vs Original Algorithm} presents the time overhead of the three algorithms on the MovieLens dataset (baseline RMSE $0.80$).
The experimental results show that the LSH-MF can achieve a $44.3X$ speedup compared to the GSM-based Top-$K$ nearest neighbourhood MF.
CULSH-MF can achieve a $196.22X$ speedup compared to the LSH-MF serial algorithm.
These results demonstrate that the proposed algorithms are efficient.

\begin{table}[htbp]
	\centering
	\setlength{\abovecaptionskip}{0pt}
	\caption{Running time (Seconds)}
	\begin{tabular}{c|cccc}
		\hline
		\hline
		Algorithm & Platform 				 & $F$   & $K$ 	 & Time   \\
		\hline
		Serial   	& Intel Xeon E5-2620 CPU & 32  & 32  & 782.64          \\
		LSH-MF      & Intel Xeon E5-2620 CPU & 32  & 32  & 17.66           \\
		CULSH-MF    & Nvidia Tesla P100 GPU  & 32  & 32  & 0.09            \\
		\hline
		\hline
	\end{tabular}
	\label{CUMF-LSH vs Original Algorithm}
\end{table}

The comparison baselines of the GSM and simLSH are set under the same experimental conditions.
To make the experiment more rigorous, a randomized control group was added, and
it randomly selects $K$ variables for each variable rather than the Top-$K$ nearest neighbours query.

	Furthermore, we compared two other LSH algorithms, random projection (RP\_cos) based on cosine distance and minHash based on Jaccard similarity.
	On sparse data, compared to the Euclidean distance, the LSH algorithms based on the cosine distance have less accuracy loss.
	In addition, minHash can approximately calculate the Jaccard similarity between sets or vectors.
	The above two LSH functions are simple and have low computational complexity,
	Furthermore, the more complex LSH functions are not suitable for high-dimensional sparse data.

\begin{figure*}[htbp]
	\centering
	\includegraphics[width=1.80in]{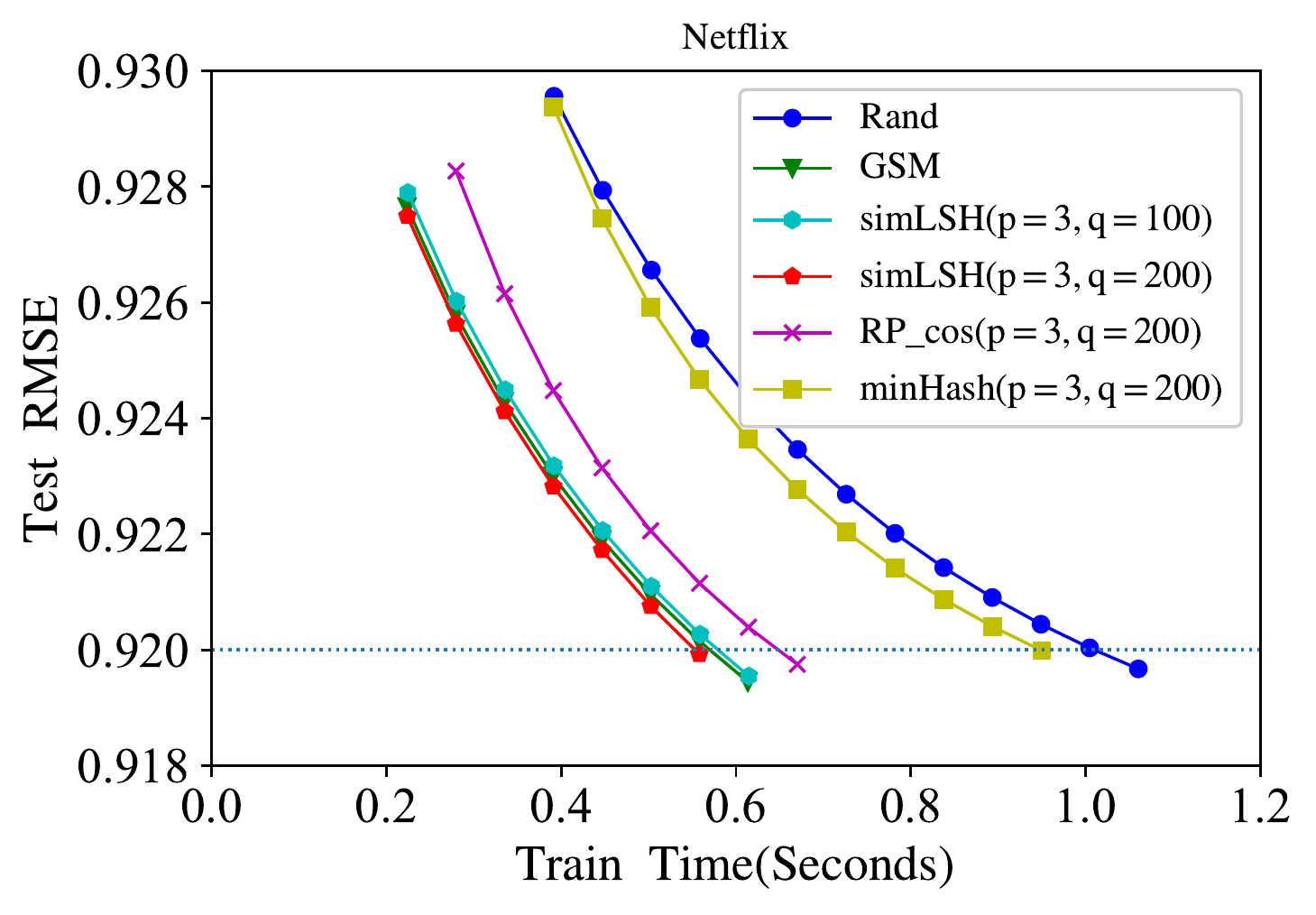}
	\includegraphics[width=1.80in]{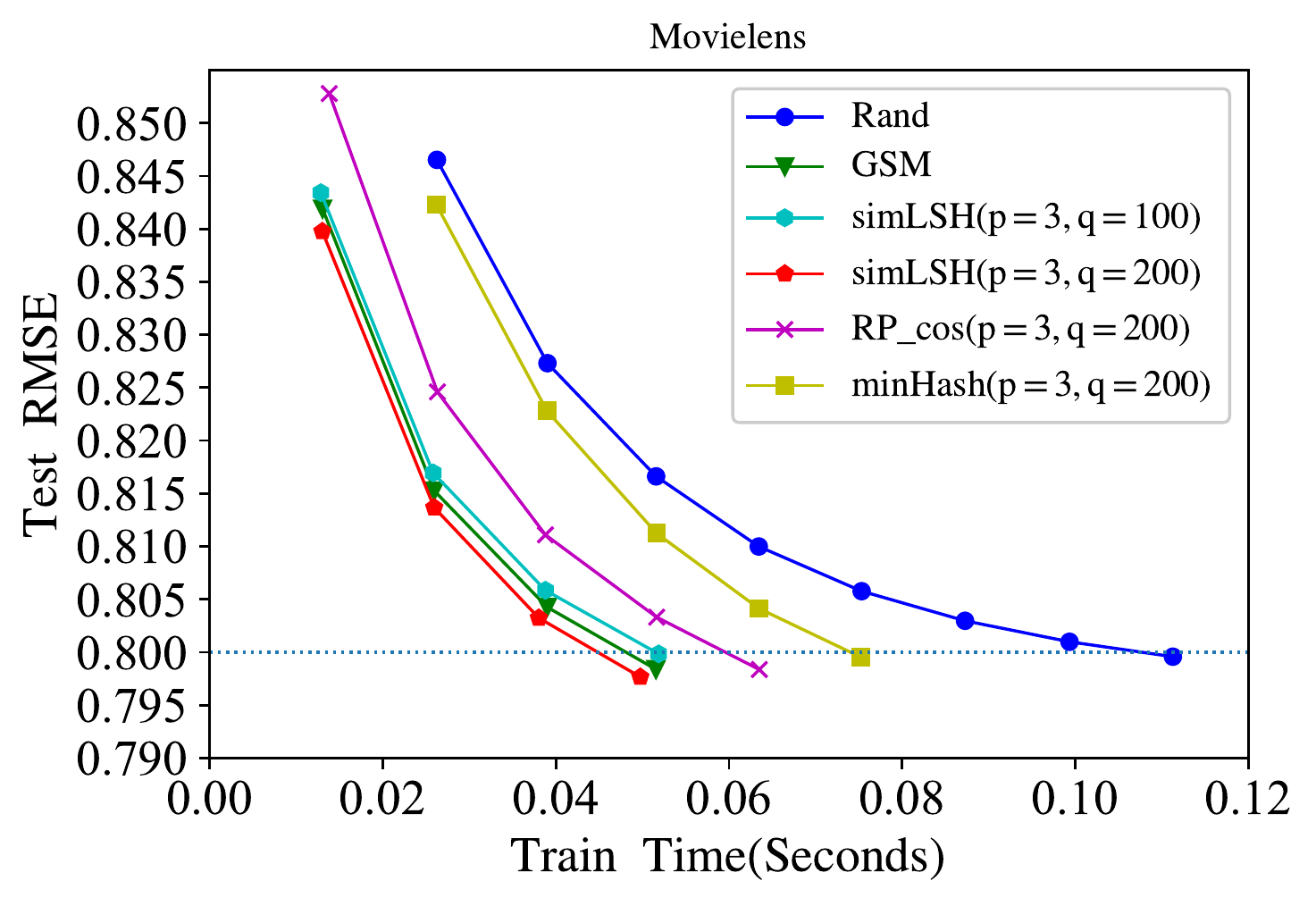}
	\includegraphics[width=1.80in]{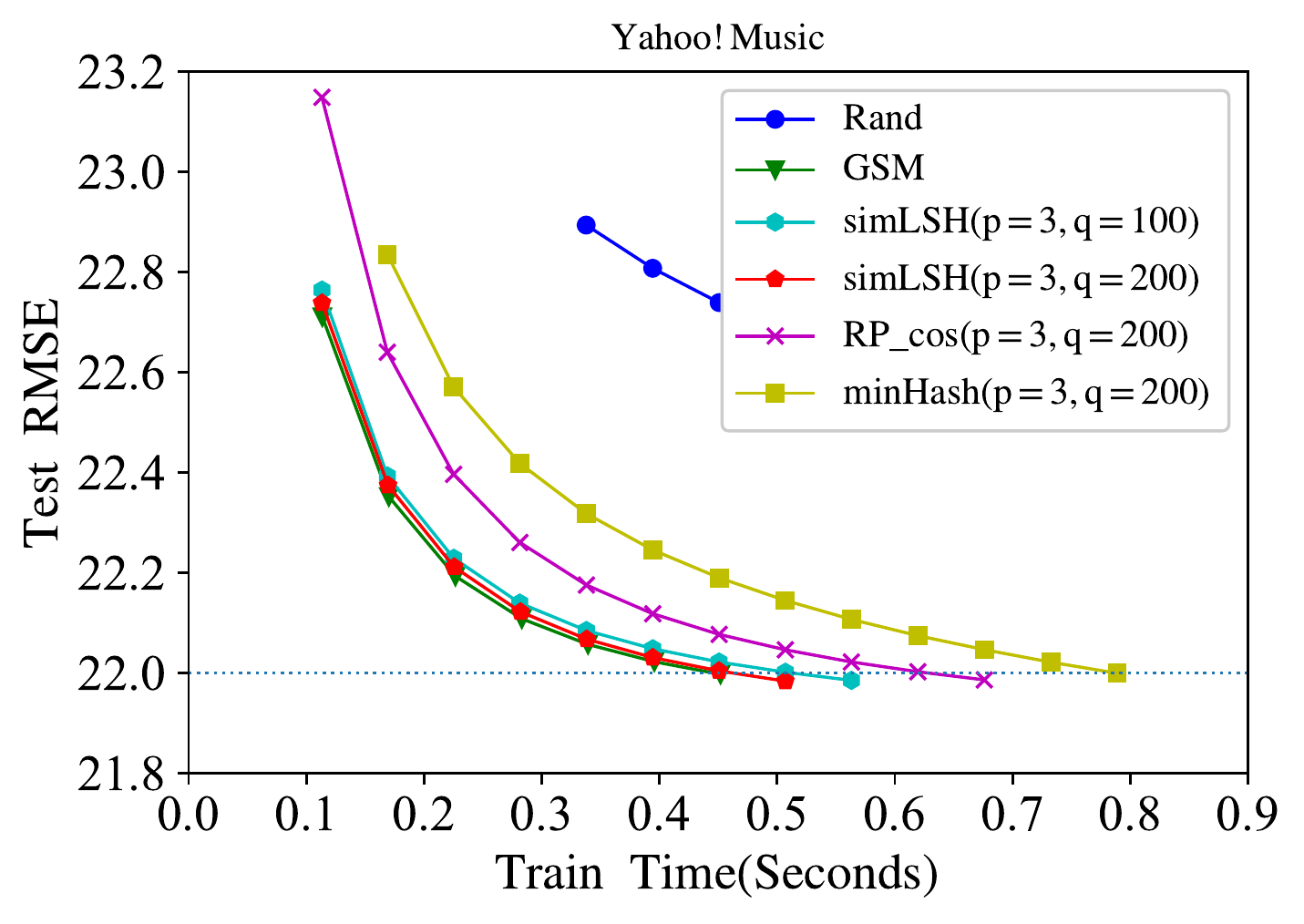}
	\caption{$RMSE$ vs time: The comparison between GSM, simLSH (various $p$ and $q$ values) and other LSH algorithms.}
	\label{lsh_culsh-mf}
\end{figure*}

The baseline $RMSE$s are $\{0.92, 0.80, 22.0\}$ for Netflix, MovieLens and Yahoo! Music, respectively.
For the MovieLens and Netflix datasets,
$\Psi(r_{i,j})=r_{i,j}^2$ is set to expand the gap between interaction values,
and the Yahoo! Music dataset has more dense interaction values.
Thus, $\Psi(r_{i,j})=r_{i,j}^4$.
We use a byte as a hash value ($G=8$) and set $\lambda_{\rho}$ as the commonly used 100.
Fig. \ref{lsh_culsh-mf} shows that the random selection method performs worse than the GSM-based method, simLSH and other LSH algorithms on the three datasets.
When the parameters $\{p, q\}$ are set as $\{p=3, q=100\}$,
simLSH is almost the same as that of the GSM.

	When the parameters $\{p, q\}$ are set as $\{p=3, q=100\}$, simLSH surpasses the GSM, and the performances of RJ\_cos and minHash are far from that of simLSH. The reason is that the datasets are very sparse, and the descent speed brought by minHash is not very impressive.

\begin{figure*}[htbp]
	\centering
	\includegraphics[width=1.80in]{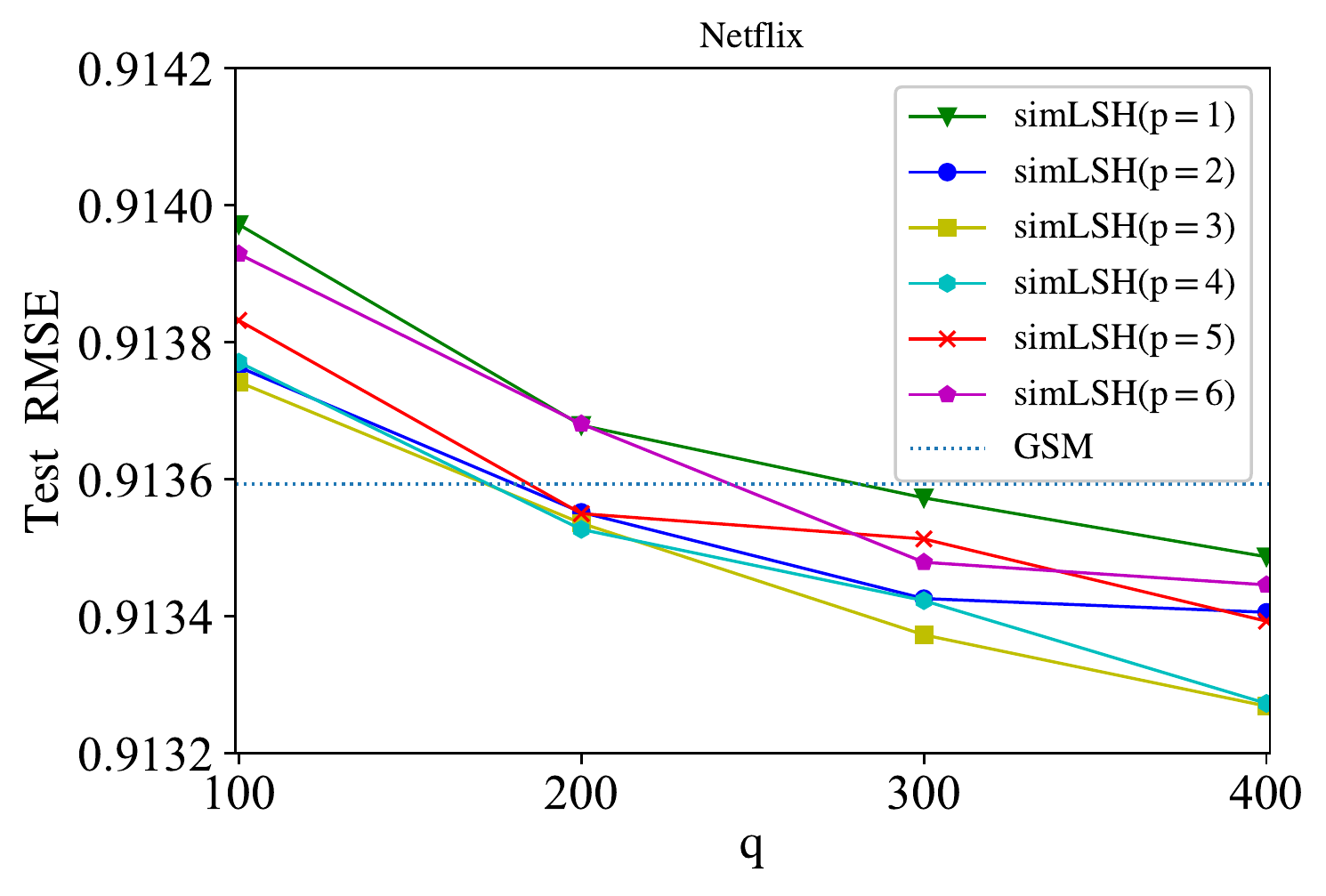}
	\includegraphics[width=1.80in]{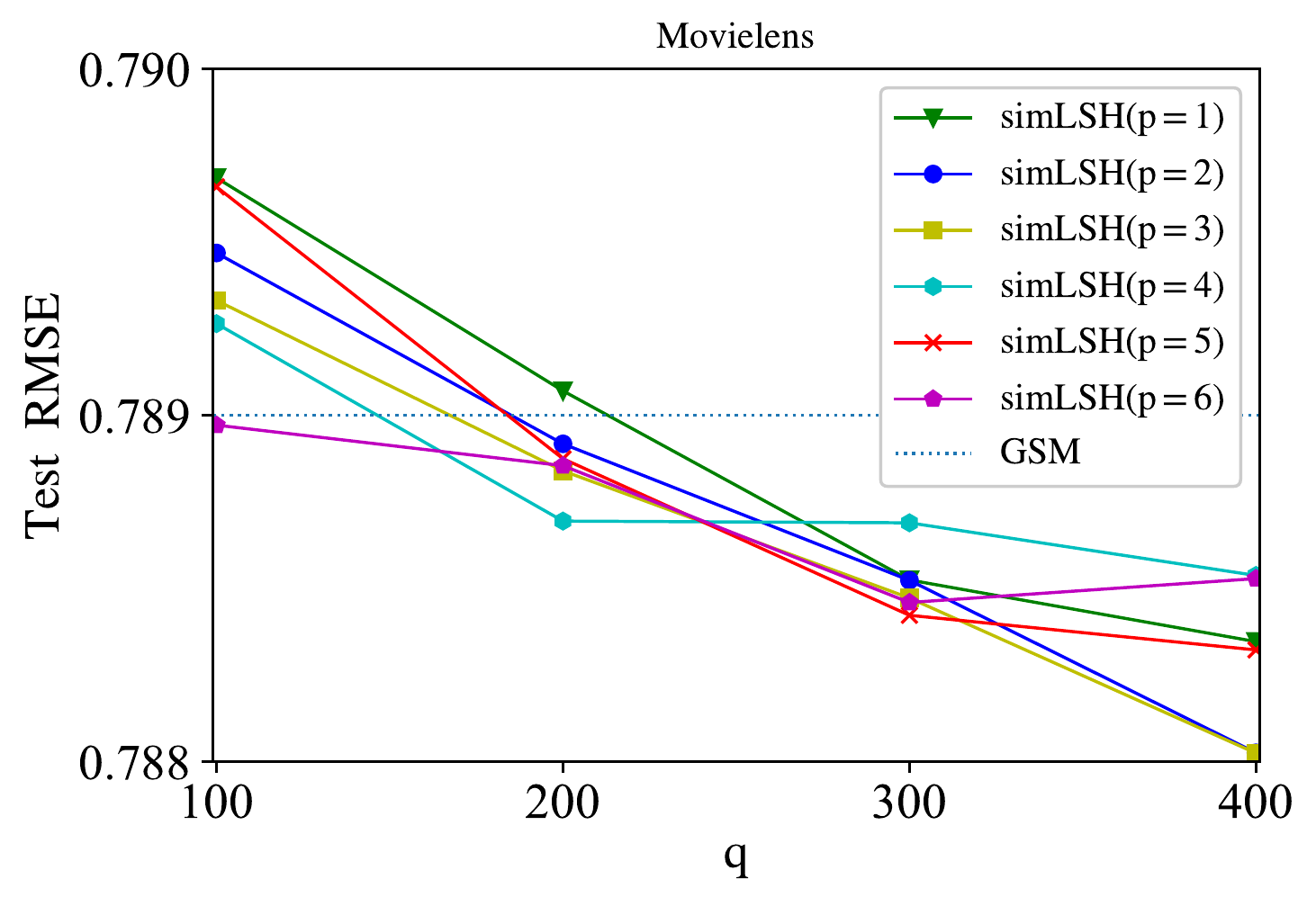}
	\includegraphics[width=1.80in]{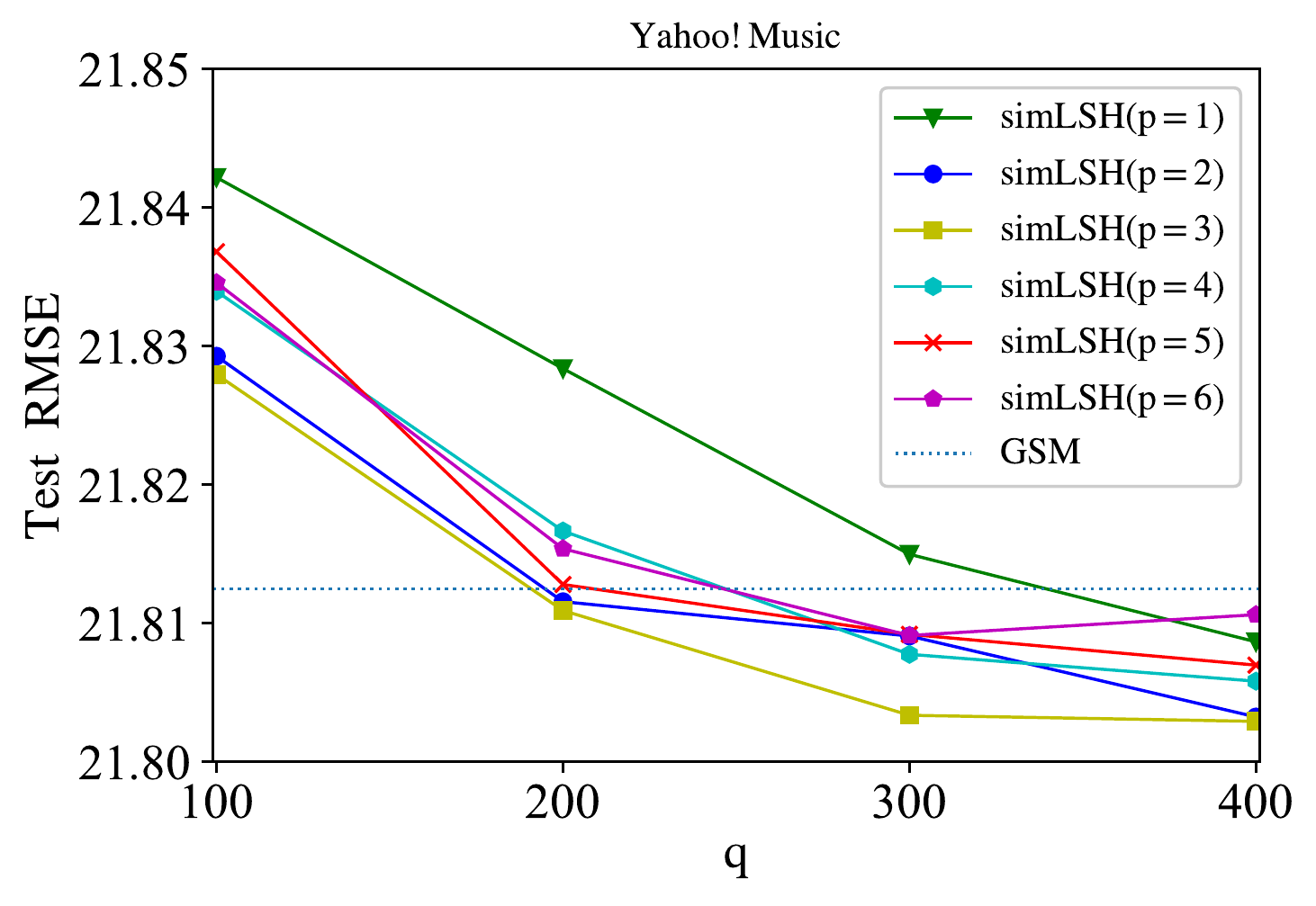}
	\caption{$RMSE$ vs influence of various value of $\{p, q\}$.}
	\label{pq_rmse}
\end{figure*}

Table \ref{lsh_time} shows the optimal $RMSE$ and the corresponding time overhead.
Table \ref{lsh_time} (top) demonstrates that simLSH can achieve a better $RMSE$ than using the GSM and
simLSH is better than the GSM and other LSH algorithms not only in descent speed but also in accuracy.
Table \ref{lsh_time} (middle) shows the time overhead of GSM, simLSH and other LSH algorithms on the three datasets, and simLSH takes much less time than the GSM.
The calculation time required for RP\_cos is slightly larger than that of simLSH,
	and minHash requires considerable calculation time due to the high dimensionality of the datasets.
Table \ref{lsh_time} (bottom) shows the spatial overhead of GSM, simLSH and other LSH algorithms on the three datasets, and simLSH takes much less space than the GSM.
Furthermore,
simLSH can surpass the GSM since it can adjust the parameters to achieve a balance between accuracy and time and
it can set appropriate parameters according to actual needs.
Fig. \ref{pq_rmse} shows the influence of various values of $\{p, q\}$ on the three datasets.
The increase in $p$ will reduce the probability of two dissimilar variables projecting to the same hash value to $P_{2}^{p}$,
but the probability $1-(1-P_1^{p})^{q}$ of two similar variables projected to the same hash value will decrease.
Choosing a suitable $p$ will achieve higher accuracy.

\begin{figure*}[htbp]
	\centering
	\includegraphics[width=1.80in]{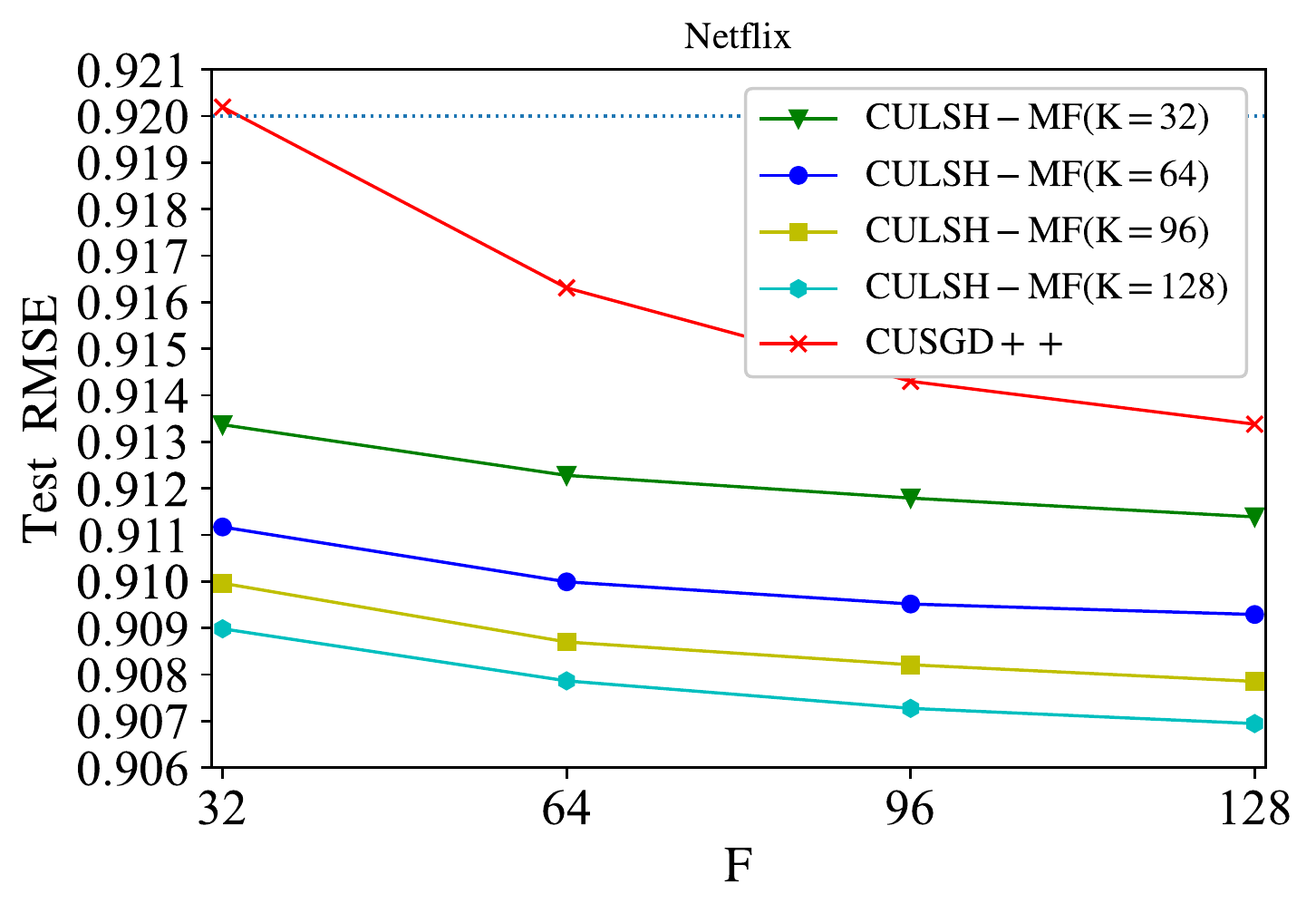}
	\includegraphics[width=1.80in]{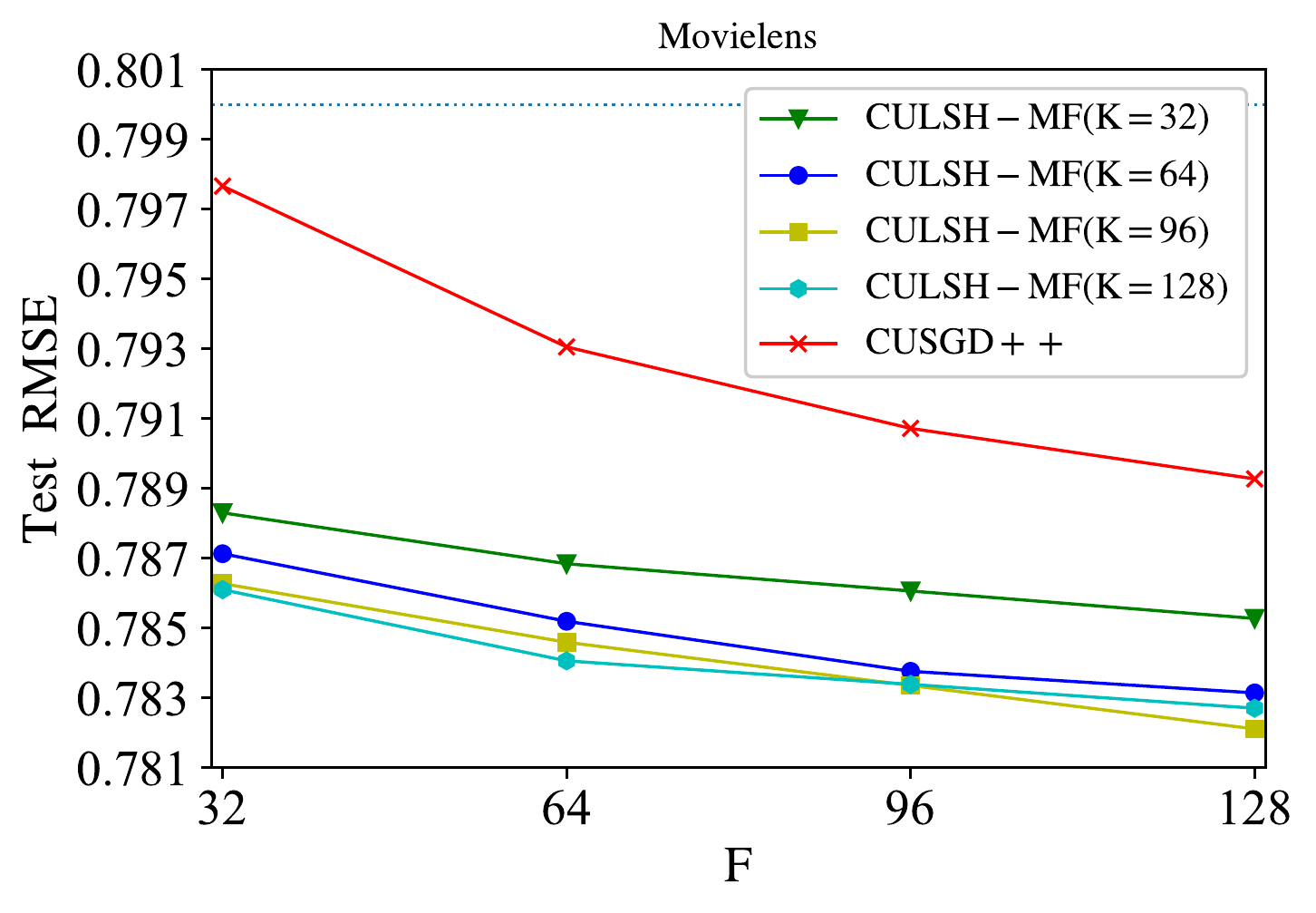}
	\includegraphics[width=1.80in]{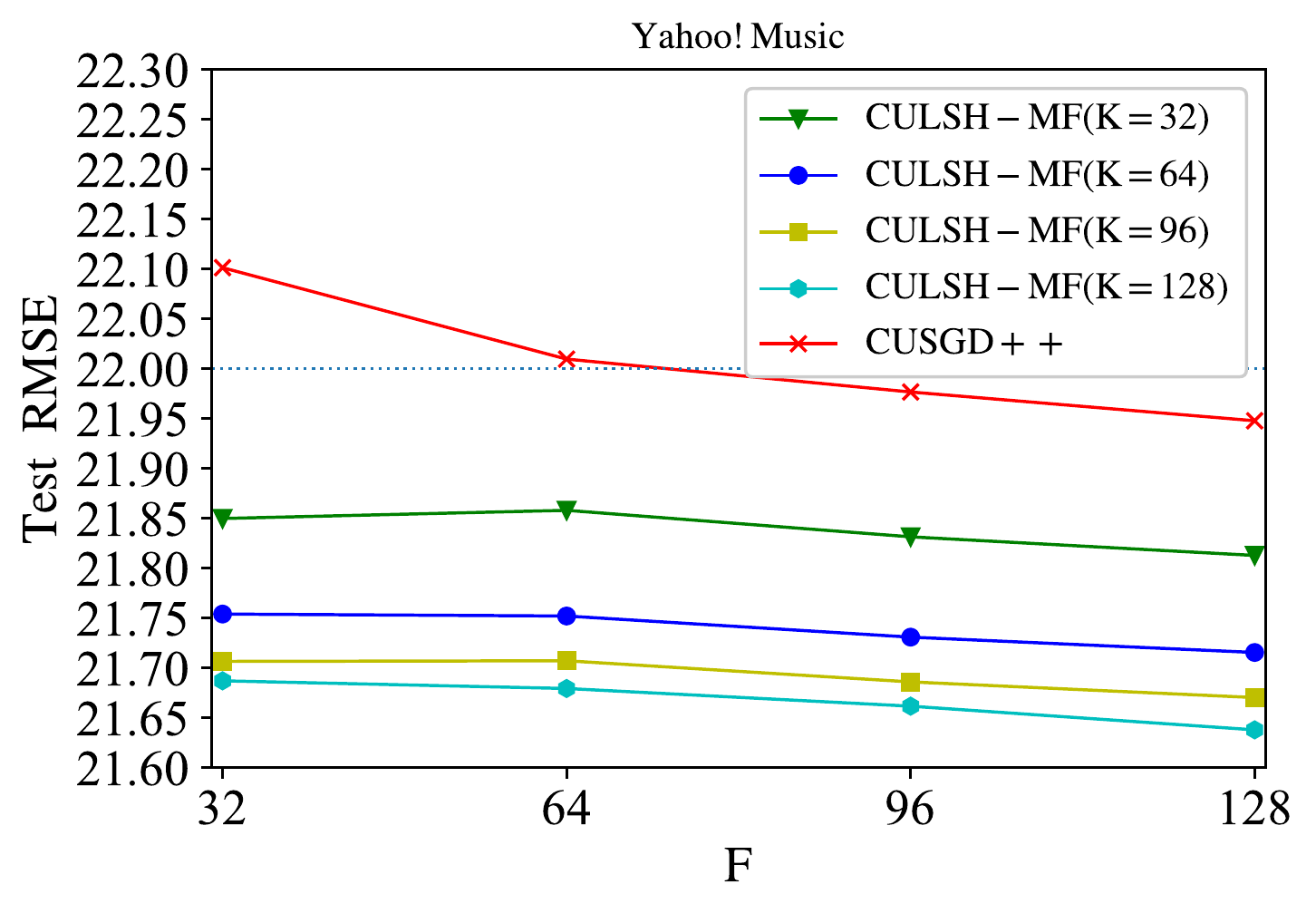}
	\caption{$RMSE$ vs influence of various value of $\{F, K\}$. Compared with $F$, increasing $K$ can reduce RMSE more.}
	\label{fk_rmse}
\end{figure*}

\begin{table}[htbp]
	\centering
	\setlength{\abovecaptionskip}{0pt}
	\caption{The optimal $RMSE$ of various Top-$K$ methods (Up), the time overhead of various Top-$K$ methods (Seconds) (Middle) and the space overhead of various Top-$K$ methods (MB) (Down)}
	\begin{tabular}{c|cccc}
		\hline
		\hline
		Indicator & Method  & Netflix  & Movielens & Yahoo! Music \\ \hline
		\multirow{6}{*}{RMSE}& Rand               & 0.9157 & 0.7947  & 21.99   \\
		&GSM                & 0.9136 & 0.7890  & 21.81   \\
		&simLSH (p=3,q=100)  & 0.9137 & 0.7893  & 21.83   \\
		&simLSH (p=3,q=200)  & 0.9135 & 0.7888  & 21.81   \\
		&RP\_cos (p=3,q=200)  & 0.9139 & 0.7896  & 21.87   \\
		&minHash (p=3,q=200)  & 0.9138 & 0.7892  & 21.82   \\ \hline 	
		\multirow{6}{*}{Time Overhead (Seconds)}&	Rand               & 0.0     & 0.0       & 0.0         \\
		&GSM                & 422.996 & 27.150    & 295.417     \\
		&simLSH (p=3,q=100)  & 15.414  & 2.777     & 25.994      \\
		&simLSH (p=3,q=200)  & 31.017  & 5.602     & 52.012      \\
		&RP\_cos (p=3,q=200)  & 47.262 & 8.184  & 78.953   \\
		&minHash (p=3,q=200)  & 270.003 & 38.224  & 319.831   \\ \hline
		\multirow{6}{*}{Space Overhead (MB)}&Rand               & 0.0     & 0.0       & 0.0         \\
		&GSM                & 1,204.578 & 434.869    & 611.209     \\
		&simLSH (p=3,q=100)  & 20.336  & 12.219     & 14.486      \\
		&simLSH (p=3,q=200)  & 40.672  & 24.438     & 28.972      \\
		&RP\_cos (p=3,q=200)  & 40.672 & 24.438 & 28.972   \\
		&minHash (p=3,q=200)  & 40.672 & 24.438  & 28.972   \\ \hline \hline	
	\end{tabular}
	\label{lsh_time}
\end{table}

\begin{figure*}[htbp]
	\centering
	\includegraphics[width=1.80in]{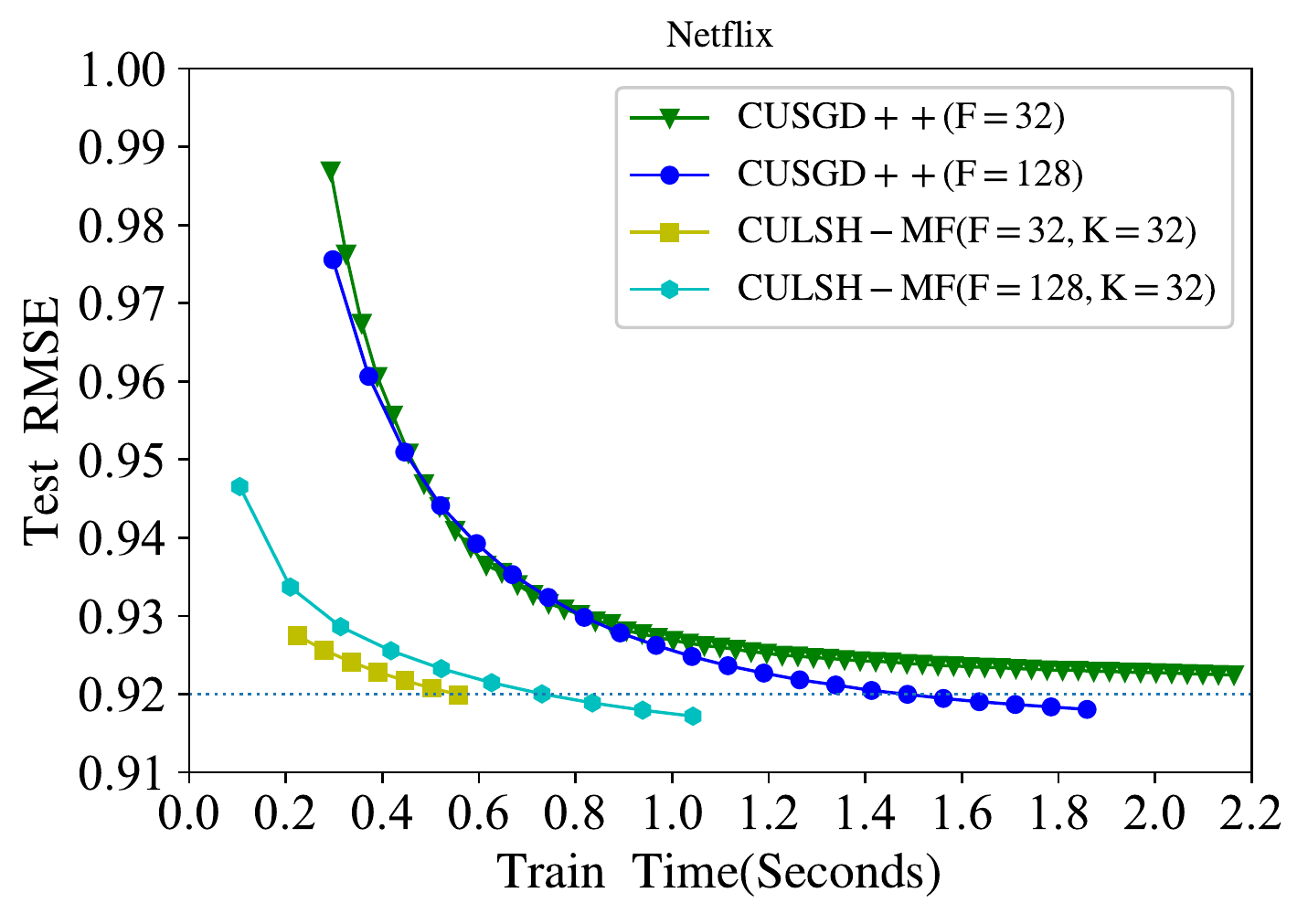}
	\includegraphics[width=1.80in]{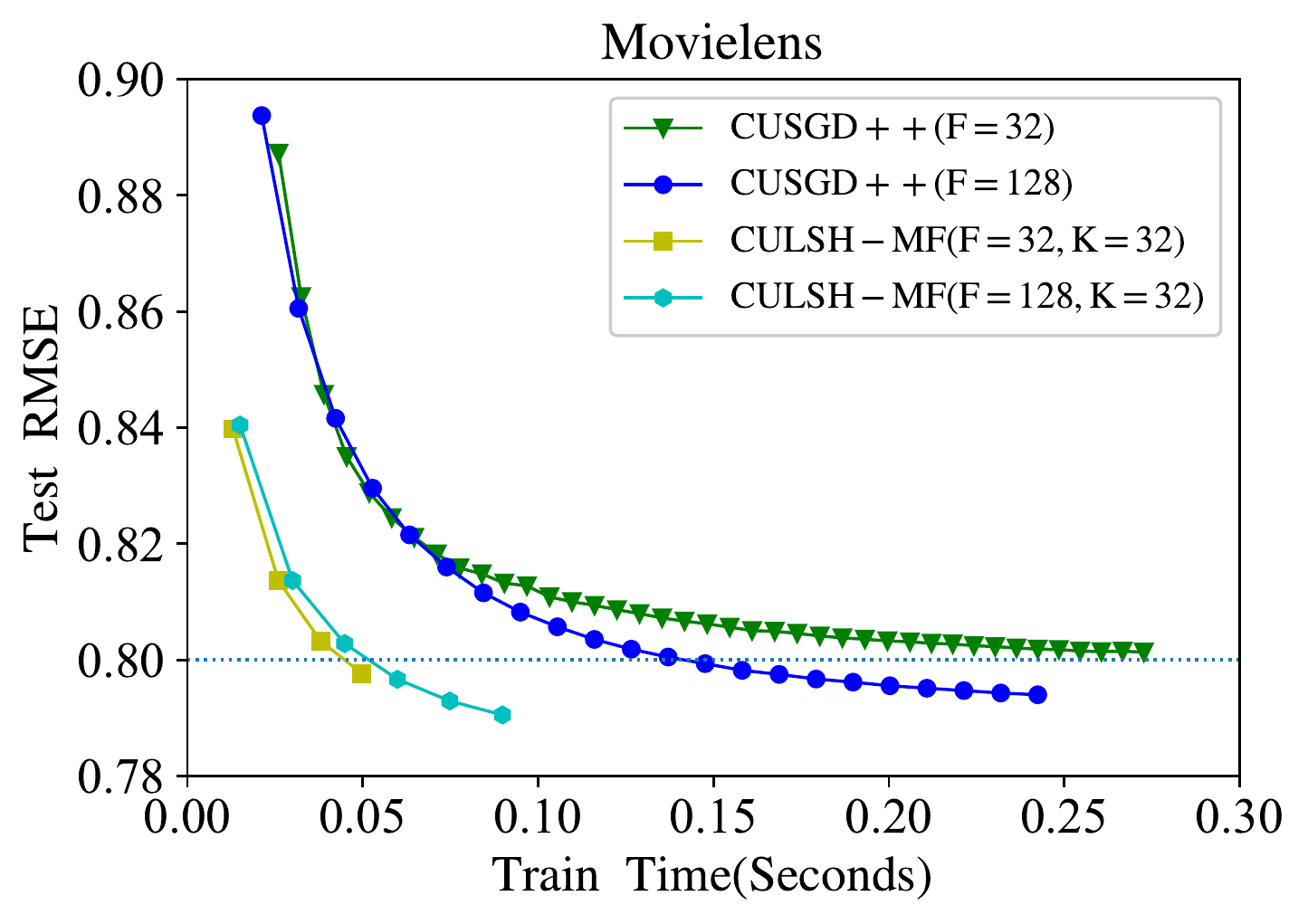}
	\includegraphics[width=1.80in]{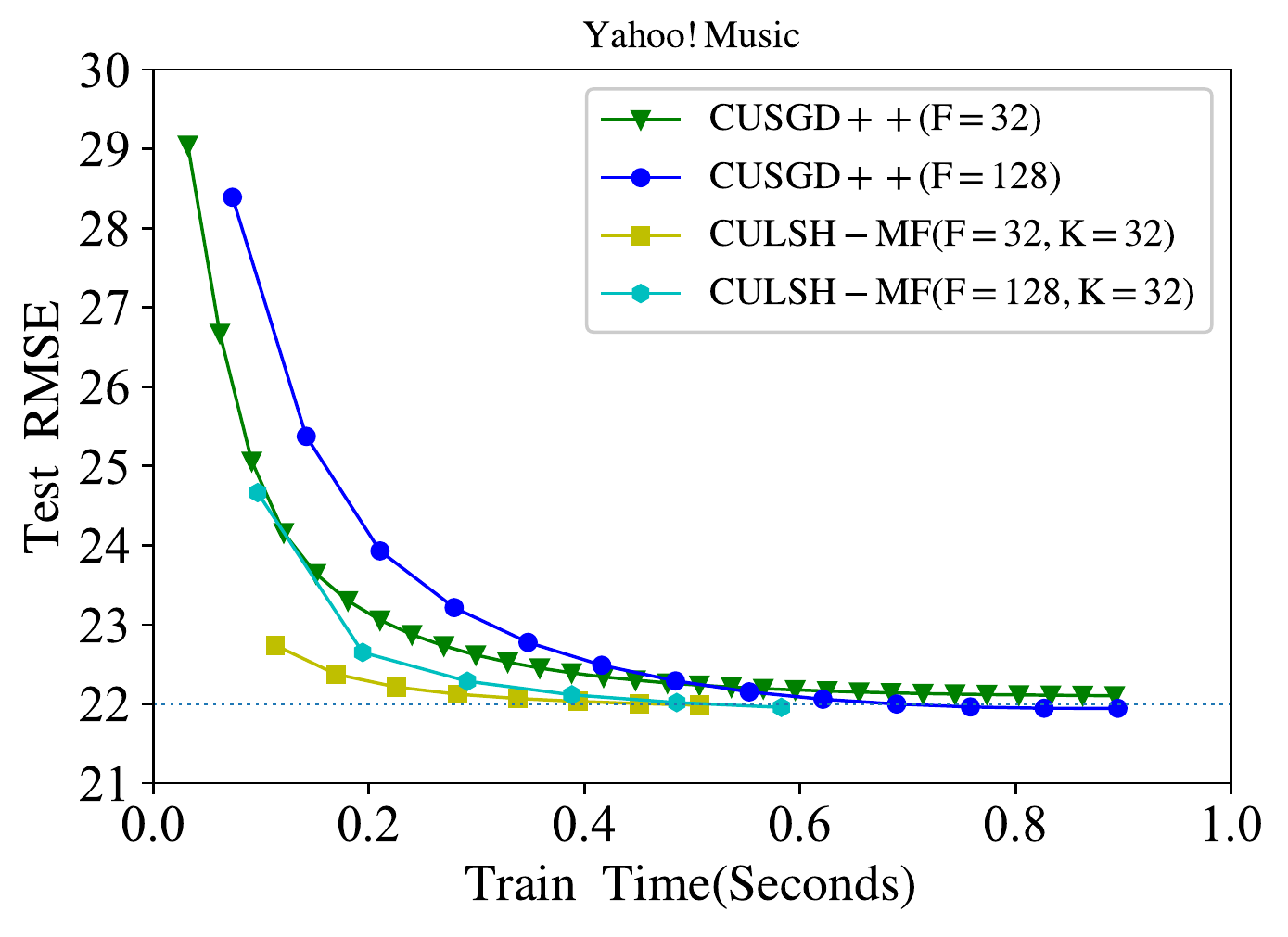}
	\caption{$RMSE$ vs time: CULSH-MF outperforms CUSGD++ on all three datasets.}
	\label{cusgd++vsculshmf}
\end{figure*}

We should select the best parameters and ensure which parameters play a greater role.
In order to ensure that the threads are fully utilized,
the parameters $\{F, K\}$ are all set as $\{32, 64, 96, 128\}$.
Fig. \ref{fk_rmse} illustrates the influences of $\{F, K\}$ on CULSH-MF.
As the Fig. \ref{fk_rmse} shows,
under the same $F$,
CULSH-MF with the neighbourhood model obtains higher accuracy than CUSGD++ without the neighbourhood model in terms of the $RMSE$.
Then, CULSH-MF is compared with CUSGD++ to demonstrate to what degree the neighbourhood model can improve the accuracy.
Fig. \ref{cusgd++vsculshmf}
shows that CULSH-MF with the parameters $\{F=128, K=32\}$ achieves a much faster descent speed than CUSGD++ with $F=128$.
The neighbourhood model with a low $K$ can greatly improve the descent speed, and it can reach the target $RMSE$ with only a few iterations.
CUSGD++ has a shorter training time per iteration, but it requires more training periods.
Thus, CULSH-MF can outperform CUSGD++ owing to the overall training time with the optimal $RMSE$.
Another noteworthy results is that CULSH-MF runs faster than CUSGD++ as the value of $F$ increases.
CULSH-MF with parameter $K=32$ can achieve $\{2.67X, 2.97X, 1.36X\}$ speedups compared to CUSGD++ when $F$ $=$ $\{32, 64, 128\}$, respectively.

\begin{table}[htbp]
	\centering
	\footnotesize
	\setlength{\abovecaptionskip}{0pt}
	\caption{$RMSE$ deviation of the noisy data and the clean data}
	\begin{tabular}{c|cccc}
		\hline \hline
		Noise Rate& Algorithm           & Netflix & Movielens & Yahoo!Music \\
		\hline
		\multirow{2}{*}{1\%}      & CUSGD++($F$=128)            & 0.00116 & 0.00157   & 0.13840     \\
		& CULSH-MF($F$=32, K=32)            & 0.00096 & 0.00166   & 0.09770     \\
		\hline
		\multirow{2}{*}{0.5\%}    & CUSGD++($F$=128)              & 0.00055 & 0.00092   & 0.06012     \\
		& CULSH-MF($F$=32, K=32)            & 0.00045 & 0.00076   & 0.04792     \\
		\hline
		\multirow{2}{*}{0.1\%}    & CUSGD++($F$=128)             & 0.00032 & 0.00040   & 0.01404     \\
		& CULSH-MF($F$=32, K=32)            & 0.00011 & 0.00006   & 0.00954     \\
		\hline
		\multirow{2}{*}{0.05\%}   & CUSGD++($F$=128)             & 0.00018 & 0.00028   & 0.00814     \\
		& CULSH-MF($F$=32, K=32)            & 0.00002 & 0.00004   & 0.00424     \\
		\hline
		\multirow{2}{*}{0.01\%}   & CUSGD++($F$=128)             & 0.00011 & 0.00016   & 0.00412     \\
		& CULSH-MF($F$=32, K=32)            & 0.00001 & 0.00002   & 0.00194     \\
		\hline
		\hline
	\end{tabular}
	\label{noise}
\end{table}

\begin{table}[htbp]
	\centering
	\footnotesize
	\setlength{\abovecaptionskip}{0pt}
	\caption{Online Data Sets}
	\begin{tabular}{c|ccc}
		\hline \hline
		Parameter & Netflix  & Movielens & Yahoo!Music \\ \hline
		$M$    & 475, 388   & 69, 180     & 580, 388      \\
		$N$    & 17, 593    & 10, 571     & 12, 532       \\
		$|\Omega|$  & 98, 339, 095 & 9, 789, 247   & 90, 752, 595  \\
		$\overline{M}$   & 4, 801  	& 698     	& 5, 862	      \\
		$\overline{N}$    & 177    	& 106     	& 126     	  \\
		$|\overline{\Omega}|$  & 733, 017   & 110, 807    & 1, 217, 617     \\ \hline \hline
	\end{tabular}
	\label{Online_Data_Sets}
\end{table}

Finally, we present the experimental results of the robustness of CULSH-MF and CUSGD++, the online learning and multiple GPU solutions of CULSH-MF.
First,
data inevitably have noise, and a robust model should suppress noise interference.
The experiment is conducted on all datasets with noise rates of $\{1\%, 0.5\%, 0.1\%, 0.05\%,$ $ 0.01\%\}$.
The experimental results in Table \ref{noise} show that
CULSH-MF has more robustness than CUSGD++, which means that the neighbourhood nonlinear model performs more robustly than the naive model.
Second,
we divide the training datasets of Netflix, MovieLens and Yahoo! Music into original set $\Omega$ and new set $\overline{\Omega}$, and $|\Omega|$ $\ll$ $|\overline{\Omega}|$.
The specific conditions of the dataset are shown in Table \ref{Online_Data_Sets}.
In the online experiments, the $RMSE$ of our online CULSH-MF on the Netflix, MovieLens, and Yahoo! Music datasets only increased by $\{0.00015, 0.00040, 0.00936\}$, respectively,
which means that online CULSH-MF avoids the retraining process.
Third, multiple GPUs can accommodate a larger data,
and CULSH-MF is extended to MCULSH-MF.
Due to the communication overhead between each GPU,
MCULSH-MF cannot reach the linear speeds, and
properly distributing communications can shorten the computation time.
CULSH-MF can obtain $\{1.6X, 2.4X, 3.2X\}$ speedups on $\{2, 3, 4\}$ GPUs, respectively,
compared to CULSH-MF on a GPU.

\begin{table}[htbp]
	\footnotesize
	\centering
	\setlength{\abovecaptionskip}{0pt}
	\caption{Time comparison (Seconds) to obtain basic HR of various nonlinear MF methods}
	\begin{tabular}{c|cc}
		\hline
		\hline
		Algorithm & Movielens1m (HR 0.65) & Pinterest (HR 0.85) \\
		\hline
		GMF      & 219.6                 & 335.1              \\
		MLP      & 940.4                 & 1289.9             \\
		NeuMF    & 308.5                 & 402.3              \\
		CULSH-MF & 0.0343                & 0.0452             \\
		\hline	
		\hline
	\end{tabular}
	\label{ncf_1}
\end{table}

%\subsection{Comparison With Deep Learning Model}
Our model also applies to recommendations for implicit feedback and has a very obvious time advantage.
NCF works well but takes too much time, and CULSH-MF can achieve similar results with a lower time overhead.
We change the loss function of CULSH-MF to the cross entropy loss function, and
the update formula will also follow the corresponding change.
This derivation is too simple and will not be repeated here.
Because the time overheads to train the deep learning models on large-scale datasets are unacceptable,
three deep learning models, e.g., Generalized Matrix Factorization (GMF), the Multilayer Perceptron (MLP) and
Neural Matrix Factorization (NeuMF), of \cite{ex152} are just tested on two small datasets, e.g., MovieLens1m and Pinterest.
1) GMF is a deep learning model based on matrix factorization
that extends classic matrix factorization.
It first performs one-hot encoding on the indexes in the sets $\{I, J\}$ of the input layer,
and the obtained embedding vectors are used as the latent factor vectors.
Then, through the neural matrix decomposition layer,
it calculated the matrix Hadamard product of factor vector $I$ and factor vector $J$.
Finally, a weight vector and the obtained vector are projected to the output layer by the dot product.
2) The MLP is used to learn the interaction between latent factor vector $I$ and latent factor vector $J$,
which can give the model greater flexibility and nonlinearity.
With the same conditions as GMF,
the MLP uses the embedded vector of the one-hot encoding of indices $I$ and
$J$ as the latent factor vector of $I$ and $J$.
The difference is that MLP concatenates latent factor vector $I$ with latent factor vector $J$.
The model uses the standard MLP; and each layer contains a weight matrix, a deviation vector, and an activation function.
3) GMF uses linear kernels to model the interaction of potential factors
while MLP uses nonlinear kernels to learn the interaction functions from data.
To consider the above two factors at the same time,
NeuMF integrates GMF and the MLP, embeds GMF and the MLP separately, and combines these two models by connecting their last hidden layers in series.
This allows the fusion model to have greater flexibility.
The Hit Ratio (HR) is used to measure the accuracy of the nonlinear models.
We use the same datasets and the same metrics.
For the same baseline HR,
we compare the time overheads of CULSH-MF and the three nonlinear models, i.e., GMF, the MLP and NeuMF.
The experimental results are shown in Table \ref{ncf_1}.
Table \ref{ncf_1} shows that the time overhead of the CULSH-MF is only $0.01\%$ that of the three nonlinear models, i.e., GMF, the MLP and NeuMF.
Furthermore, the parameters of the CULSH-MF are much smaller than those of the three nonlinear models, i.e., GMF, the MLP and NeuMF.

\section*{Acknowledgment}
The research was partially funded by the National Key R\&D Program of China (Grant No. 2020YFB2104000) and the Programs of National Natural Science Foundation of China (Grant Nos. 61860206011, 62172157).
This work has been partly funded by the Swiss National
Science Foundation NRP75 project (Grant No. 407540\_167266) and
the China Scholarship Council (CSC) (Grant No. CSC201906130109).

%%
%% The acknowledgments section is defined using the "acks" environment
%% (and NOT an unnumbered section). This ensures the proper
%% identification of the section in the article metadata, and the
%% consistent spelling of the heading.

%%
%% The next two lines define the bibliography style to be used, and
%% the bibliography file.
\bibliographystyle{ACM-Reference-Format}
\bibliography{sample-base}

%%
%% If your work has an appendix, this is the place to put it.

\end{document}